\documentclass[a4paper, 11pt]{article}
\usepackage[pdftex, a4paper]{geometry}
\usepackage{graphicx}			

\newcommand*\mcap{\mathbin{\mathpalette\mcapinn\relax}}
\newcommand*\mcapinn[2]{\vcenter{\hbox{$\mathsurround=0pt
  \ifx\displaystyle#1\textstyle\else#1\fi\bigcap$}}}

\newcommand*\mcup{\mathbin{\mathpalette\mcupinn\relax}}
\newcommand*\mcupinn[2]{\vcenter{\hbox{$\mathsurround=0pt
  \ifx\displaystyle#1\textstyle\else#1\fi\bigcup$}}}

\DeclareFontFamily{OT1}{pzc}{}
\DeclareFontShape{OT1}{pzc}{m}{it}{<-> s * [1.200] pzcmi7t}{}
\DeclareMathAlphabet{\mathpzc}{OT1}{pzc}{m}{it}
			
\usepackage{fancyhdr}
\usepackage{amsmath}
\usepackage{amsfonts}
\usepackage{dsfont}
\usepackage{mathrsfs}
\usepackage{amssymb}
\usepackage{bbm}
\usepackage{epsfig}

\usepackage[english]{babel}
\usepackage[all]{xy}

\def\T{^{\rm\tiny T}}

\newtheorem{theorem}{Theorem}
\newtheorem{definition}{Definition}
\newtheorem{lemma}{Lemma}
\newtheorem{remark}{Remark}
\newtheorem{proposition}{Proposition}

\title{\bf Consensus  over Random Graph Processes:  Network  Borel-Cantelli Lemmas for  Almost Sure Convergence
}
\date{}

\author{Guodong Shi, Brian D. O. Anderson, and Karl Henrik Johansson
}

\begin{document}

\maketitle
\begin{abstract}
Distributed  consensus computation over  random graph processes is considered.   The  random graph process is defined   as a sequence of random variables which  take values from the set of all possible digraphs over the  node set.  At each time step,  every node updates its state based on a Bernoulli trial,  independent in time and among different nodes: either  averaging  among the neighbor set generated  by the random graph,  or sticking with its current state.
   Connectivity-independence  and arc-independence  are introduced to capture the fundamental influence of the random graphs on the  consensus convergence.  Necessary and/or sufficient conditions are presented on the success probabilities of the Bernoulli trials  for the network to reach a global almost sure consensus, with some sharp threshold established revealing a consensus zero-one law.    Convergence rates are established by lower and upper bounds of  the $\epsilon$-computation time. We also generalize the concepts of connectivity/arc independence to their analogues from the $*$-mixing point of view, so that our results apply to a very wide class of graphical models, including  the majority of random graph models in the literature, e.g., Erd\H{o}s-R\'{e}nyi, gossiping,  and Markovian random graphs. We show that under  $*$-mixing, our convergence  analysis continues to hold and the corresponding  almost sure consensus conditions are established. Finally, we further investigate almost sure finite-time convergence of random gossiping algorithms, and prove that  the Bernoulli trials  play a key role in ensuring finite-time convergence. These results add to  the understanding of the interplay between random graphs, random computations, and convergence probability for distributed information processing.


\end{abstract}

\section{Introduction}
In recent years, there has been considerable research interest on distributed algorithms for  information exchange, estimation, and  computation over networks. Such algorithms have  a variety of potential applications in sensor networks, peer-to-peer networks, wireless networks and networked control systems. Targeting the design of simple decentralized algorithms for computation or estimation,  where nodes exchange information only with their neighbors, a distributed consensus algorithm  serves as a primitive towards more sophisticated information processing algorithms, e.g.,  \cite{jad03, boyd, shah, roy}.

The investigation of the consensus problem has a long history in several scientific fields including computer science \cite{cs2,cs3}, engineering \cite{mar, ren}, and social science \cite{degroot,daron,como}. Deterministic consensus algorithms have been extensively studied for both time-invariant and time-varying communication graphs in the literature, with efforts typically devoted to finding  proper connectivity  conditions  that can ensure a desired collective convergence  \cite{saber04, fax, mor,  caoming1}. These efforts can indeed be traced back to the study of ergodicity of non-homogeneous Markov chains \cite{haj,wolf} and distributed decision making \cite{tsi}.  In addition, researchers were also interested in the design of weighted averaging algorithms to reach a faster consensus, or to  reach a consensus with asynchronous computations \cite{boyd1,  lifted}.

On the other hand, the theory of random graphs \cite{bollobas} is fundamental for the study of in-network  communication, information processing, and opinion aggregation, e.g., \cite{Ramam2005,cap2,daron}.     Consensus algorithms that are carried out over a random graph process have also drawn attention.   In \cite{hatano}, the authors studied linear
consensus dynamics with communication graphs defined as a sequence of independent, identically distributed (i.i.d.) Erd\H{o}s--R\'{e}nyi random graphs, and almost sure convergence was shown.  Then in \cite{wu}, the analysis was generalized to directed Erd\H{o}s-R\'{e}nyi  graphs. Mean-square performance for consensus algorithms over i.i.d. random graphs was studied in \cite{fagnani1}, and the influence of random packet drop was investigated in \cite{fagnani2}.  In \cite{jad2},  a necessary
and sufficient condition was presented   on almost sure asymptotic consensus for i.i.d. models.  In \cite{moura3,moura1,bamieh}, the authors studied  distributed average
consensus in sensor networks with quantized data and random link failures.   In \cite{markov}, the communication graph was described as a finite-state Markov chain  where each graph corresponds to one state of the chain, and almost sure consensus was concluded  by investigating the connectivity of the closed positive recurrent sets of the Markovian random graph. In \cite{aysal},  convergence to consensus was studied under more general  linear consensus algorithms, where the random update and control matrices were determined by  possibly non-stationary stochastic matrix processes coupled with disturbances. In \cite{Tahbaz-Salehi2010}, a general model was investigated for  consensus computations over stationary random graph processes with a necessary and sufficient condition presented on the expected graphs regarding almost sure consensus.  It is worth emphasizing that besides convergence to a (not necessarily average) consensus, the focus of  \cite{fagnani1} (and later \cite{daron}) was on the error with respect to the average of the initial values, which is far more challenging.

Classical random graph theory suggests that many important properties of static random  graphs appear suddenly as some  parameter  is smoothly adjusted from a probabilistic point of view \cite{bollobas}. To be precise, there exists usually a function,  called a  threshold, of a parameter (e.g., the graph size), and  a quantity (e.g., probability of edge appearance) such that a given behaviour appears asymptotically in the parameter if and only if
such a quantity grows faster than the threshold. The phenomenon is called a zero-one law. Pioneered in \cite{er}, these sharp probabilistic phase transitions have been a central motif  in the study of random graphs \cite{cap2,rg1,rg2,rg3}. In fact, as early as  in \cite{jad2},   a zero-one law  has been shown to exist  for consensus convergence over i.i.d. random graph processes.  Naturally, one may wonder,  with  possible  unwillingness or inability for a node to update,  would there be any threshold condition which leads to a similar zero-one law with respect to collective convergence of consensus dynamics over random graphs? Although various  results have been established to guarantee probabilistic consensus  \cite{hatano,wu,fagnani1,fagnani2,aysal}, the literature still lacks  a general model and analysis for consensus dynamics over random graphs which can accurately describe the fundamental influence  of graph processes and whether  such a zero-one law could arise.

To this end,  in this paper, we study  the almost sure convergence of a randomized consensus algorithm over random graph processes.  A general random graph process is adopted  to model node interactions  as a sequence of random variables which take values from the set of all possible graphs with the given node set. Another random node update process is independently built upon this random graph process, in that  at each time step, every node independently updates its state as a weighted average of its neighbors' states or sticks with its current state.  The choice is a  Bernoulli trial with success probability $P_k$.  We introduce connectivity-independence and arc-independence for the considered random graph processes. For connectivity-independent graphs, we show that $\sum_k P_k^{n-1} =\infty$ is a sufficient condition for almost sure consensus, where $n$ is the number of nodes. For arc-independent graphs, we show that $\sum_k P_k =\infty$ is a sharp threshold, i.e., the consensus probability is zero for almost all initial conditions when the sum converges, while it is one for all initial conditions when  the sum diverges. In other words, a zero-one law is established for the presented randomized consensus processing over arc-independent graphs, and we see that the success probability of the node updates defines the parameter for the corresponding threshold function. Hence, also consensus computations over random graph processes show the sudden transition similar to Erd\H{o}s--R\'{e}nyi graphs, and other models in the literature \cite{bollobas}.

\subsection{Problem Definition}
 A directed graph (digraph)  over $\mathrm{V}$ is defined as
$\mathcal{G}=(\mathrm{V}, \mathcal{E})$, where nodes are indexed in the set  $\mathrm {V}=\{1,2,\dots,n\}$ and $\mathcal{E}$ is the set of {\it arcs}, i.e.,    ordered pairs of distinct  nodes \cite{god}.  An arc  from $i$ to $j$ is denoted as $(i,j)$. There are  $2^{n(n-1)}$ different digraphs with node set $\mathrm {V}$. We label these graphs from $1$ to $2^{n(n-1)}$ by an arbitrary order.  In the following, we will identify an integer in $\mathscr{G}=\{1,\dots,2^{n(n-1)}\}$ with the corresponding graph in this order. Time is slotted at $k=0,1,\dots$. We denote by $\mathrm{G}=(\mathrm {V},\mathrm
{E})$ a random graph, which by definition  is   a random variable (over some underlying probability space)  that takes value in $\mathscr{G}$.

Let   $\mathrm{G}_k=(\mathrm {V},\mathrm
{E}_k), k=0,1,\dots$ be  a sequence of random graphs over node set $\mathrm{V}$.
We call node $j$ a {\it neighbor} of $i$ at time $k$ if  $(j,i)\in \mathrm{G}_k$.
 Each node is supposed to always be a neighbor of itself. Denote $\mathrm{N}_i(k)$ as the set of neighbors  of node $i$ at time $k$.
  Each node $i$ holds a state $x_i(k)\in\mathbb{R}$ at time $k$. Let $\langle P_k \rangle$ be a given deterministic sequence with $0< P_k<1$ for all $k$.
  Independent of the graph process, node states, and other nodes, the updating rule of node $x_i(k)$ is as follows:
\begin{equation}\label{9}
	x_i(k+1) =
	\begin{cases}
		\sum_{j\in \mathrm{N}_i(k)}a_{ij}(k)x_j(k), & \text{with probability $P_k$};\\
		\ \ x_i(k), & \text{with probability $1-P_k$}.\\
	\end{cases}
\end{equation}
Here $a_{ij}(k)$ represents the weight of arc $(j,i)$. We use the following weights rule as our standing assumption, cf., \cite{jad03,caoming1,tsi2}.

\vspace{2mm}
\noindent{\bf Assumption} {\it (Weights Rule)} (i) $\sum\limits_{j\in \mathrm{N}_i(k)}a_{ij}(k)=1$ for all $i$ and $k$; (ii)   there exists a constant $\eta>0$ such that $a_{ij}(k)\geq \eta$ for all $i$, $j$ and $k$.
\vspace{2mm}

Algorithm (\ref{9}) can   also be written as the following dynamics:
\begin{align}
x_i(k+1) = \chi_i(k) \Big( \sum_{j\in \mathrm{N}_i(k)}a_{ij}(k)x_j(k) \Big)+\big(1-\chi_i(k)\big) x_i(k),
\end{align}
where  $\chi_i(k), \ i=1,\dots,n,\  k\in\mathbb{N}$ are independent Bernoulli random variables  with $\mathbf{E}[\chi_i(k)]=P_k$ for all $i$, which are  also independent with $\langle \mathrm{G}_k \rangle$.

The randomized information processing (\ref{9}) describes the possible  unwillingness or inability of a node to update, even if it receives information from other nodes.  For instance, in the opinion dynamics of  social networks, forceful belief exchanges may happen randomly by Bernoulli decisions for individuals so that misinformation  may be spread \cite{daron}.  From an engineering viewpoint, in wireless communication nodes may be asleep or broken randomly due to the unpredictability of the environment and the unreliability of the networked communication \cite{fagnani1,moura1}. The above standing assumption is adopted throughout the paper without specific mentioning.  We note that in order to satisfy  the standing assumption,  the $a_{ij}(k)$ can  be either  deterministic  or random, e.g.,  $a_{ij}(k)={1}/{n}$ for $j\neq i\in\mathrm{N}_i(k)$ and $a_{ii}(k)=1-|\mathrm{N}_i(k)|/n$; or $a_{ij}(k)=1/|\mathrm{N}_i(k)|$ for all $j\in\mathrm{N}_i(k)$.

 Our interest is in the convergence of the  randomized consensus algorithm and  the time it takes for the network to reach a consensus. Let
$
x(k)=(x_1(k)\dots x_n(k))\T\in \mathbb{R}^{n}
$ be the random sequence driven by the randomized algorithm (\ref{9}) for initial condition $x^0=(x_1(0) \dots x_n(0))\T$. Denote
$$
 H(k):=\max_{i\in \mathrm{V}} \big\{x_{i}(k)\big\}, \quad h(k):=\min_{i\in \mathrm{V}}\big\{x_{i}(k)\big\}
$$
as the maximal and minimal   states among all nodes, respectively, and define $\mathcal{H}(k):= H(k)-h(k)$ as a consensus measure. Let $\mathbf{P}$ be the probability measure capturing all the randomness in
$\big\langle x(k)\big \rangle$, and whenever necessary we write $\mathbf{P}_{x^0}$ indicating the probability measure for initial value $x^0$.

Under our standing assumption there always holds that  $\mathcal{H}(k+1)\leq \mathcal{H}(k)$.  We introduce the following definition.
\begin{definition} Global almost sure (a.s.) {\it consensus}   is achieved for Algorithm  (\ref{9})    if $\mathbf{P}\big(\lim_{k\rightarrow \infty} \mathcal{H}(k)=0\big)=1$
for all $x^0 \in \mathbb{R}^{n}$. Moreover, for any $0< \epsilon<1$, the $\epsilon$-computation time  is defined as
\begin{equation*}
\mathscr{T}_{\rm com}(\epsilon)\doteq\sup_{x^0\in \mathbb{R}^n} \inf \Big\{ k:\ \  \mathbf{P}\big({\mathcal{H}(k)}\geq \epsilon {\mathcal{H}(0)}\big)\leq \epsilon\Big\}.
\end{equation*}
\end{definition}

\subsection{Main Results}
Let $x^0$ be considered under the standard Lebesgue measure over $\mathbb{R}^n$. We first present the following  result on the impossibility of consensus,
 which is built for arbitrary random graph process
$\langle \mathrm{G}_k\rangle$.

\begin{theorem}\label{theorem-Impossibility} (i). Global a.s. consensus can be achieved for
Algorithm (\ref{9}) only if $\sum_{k=0}^\infty P_k=\infty$, and a universal  lower bound for $\mathscr{T}_{\rm com}(\epsilon)$ is given by
$$
\mathscr{T}_{\rm com}(\epsilon)\geq \sup\Big\{ k: \ \ \sum_{i=0}^{k-1}\log (1-P_i)^{-1}\leq \frac{\log \epsilon^{-1}}{2}\Big\}.
$$

\noindent (ii). Suppose $\sum_{k=0}^\infty P_k<\infty$ and $\big\{a_{ij}(k):i,j\in\mathrm{V}, k\in\mathbb{N}\big\}$ contains at most countably many elements. Then one of the following must happen:
\begin{itemize}
\item[a)] there exists a constant $p^\flat>0$ such that $
\mathbf{P}_{x^0}\big(\lim_{k\rightarrow \infty} \mathcal{H}(k)=0\big)\geq p^\flat
$
for all $x^0 \in \mathbb{R}^n$.

\item[b)] $
\mathbf{P}_{x^0}\big(\lim_{k\rightarrow \infty} \mathcal{H}(k)=0\big)=0
$
for almost all  $x^0 \in \mathbb{R}^n$.
\end{itemize}

\noindent (iii). Suppose $\sum_{k=0}^\infty P_k<\infty$ and   there exists a constant $a_\ast$ such that $a_{ii}(k)\geq a_\ast>1/2$ for all $i$ and $k$.  Then
$\mathbf{P}\big(\lim_{k\rightarrow \infty} \mathcal{H}(k)=0\big)=0$ for almost all  $x^0 \in \mathbb{R}^n$.
\end{theorem}

Recall that a digraph $\mathrm {G}$ is said to be {\it
quasi-strongly connected}  if $\mathrm {G}$ has a directed spanning tree \cite{ber}. We introduce the following definition.
\begin{definition}
 The random graph process $\langle \mathrm{G}_k\rangle$ is called   connectivity-independent if the events
$$
\big\{\mathrm {G}_k\mbox{ is quasi-strongly connected}\big\},k\in\mathbb{N}
$$
are mutually independent.
\end{definition}

For connectivity-independent graphs, we present the following almost sure convergence result.
\begin{theorem}\label{theorem-connectivity-independent}
Suppose $ \langle\mathrm {G}_{k}\rangle$ is connectivity-independent and there exists a constant $0<q<1$ such that
$ \mathbf{P} \big( \mathrm {G}_{k }\mbox{ is quasi-strongly connected}\big) \geq q$ for all $k$. Assume  in addition that $P_{k+1}\leq P_k$.
Then Algorithm  (\ref{9}) achieves global a.s. consensus  if $\sum_{k=0}^\infty {P}_k^{n-1} =\infty$, and  an upper bound of  $\mathscr{T}_{\rm com}(\epsilon)$  is given as
$$
\mathscr{T}_{\rm com}(\epsilon)\leq \inf\Big\{M:\ \  \sum_{i=0}^{M-1}\log \Big(1-\frac{(q\eta)^{(n-1)^2}}{2}\cdot {P}_{(i+1)(n-1)^2}^{n-1}\Big)^{-1}\geq {\log \epsilon^{-2}}\Big\}\times (n-1)^2.
$$
\end{theorem}

 We believe that the convergence condition given in Theorem \ref{theorem-connectivity-independent} is  reasonably  tight since the probability that all the arcs in Algorithm (\ref{9}) are active at time $k$ is $P_k^n$,  while one single inactive arc may be enough to break the connectivity of the  graph (note that even the condition  $\sum_{k=0}^\infty {P}_k^{n-1} =\infty$ allows $P_k$ to  decrease  to zero). The condition $P_{k+1}\leq P_k$ is a technical assumption which enforces certain regularity of the node update frequencies\footnote{The condition $P_{k+1}\leq P_k$ is also to make the problem more interesting since the main challenge is to reach consensus with as small $P_k$ as possible. Certainly the counter  condition $P_{k+1}\geq P_k$ will ensure convergence  straightforwardly.}. While the sufficient convergence condition $\sum_{k=0}^\infty {P}_k^{n-1} =\infty$ indicates  it is more difficult to reach  consensus over large networks.

Nevertheless, connectivity  is a global property of a graph, and  indeed it does not rely on any specific arc. The next definition is on the independence of the existence of the arcs in the graph process.
\begin{definition}
 Let $\mathcal{G}^\dag=(\mathrm{V},\mathcal{E}^\dag)$  be a (deterministic) digraph with $\mathcal{E}^\dag=\big\{(i_\tau,j_\tau):\tau=1,\dots, |\mathcal{E}^\dag|\big\}$.
Denote $\mathpzc{E}_k(\tau):=\big\{(i_\tau,j_\tau)\in\mathrm {G}_k\big\}$ for $\tau=1,\dots, |\mathcal{E}^\dag|$ and $k\in \mathbb{N}$. Then $\langle \mathrm{G}_k\rangle$ is called  arc-independent with respect to $\mathcal{G}^\dag$ if the events
$$
\mathpzc{E}_k(\tau_k), k\in \mathbb{N}
$$
 are mutually  independent for any deterministic sequence  $\langle \tau_k \rangle $ with each $\tau_k$ taking value from $\{1,\dots, |\mathcal{E}^\dag|\}$. We call $\mathcal{G}^\dag$  a basic graph of the random graph process $\langle \mathrm{G}_k\rangle$.
\end{definition}

For arc-independent graphs, the following result holds.
\begin{theorem}\label{theorem-ARC-independent}
Suppose $ \langle\mathrm {G}_{k}\rangle$ is arc-independent  with respect to the basic graph $\mathcal{G}^\dag=(\mathrm{V},\mathcal{E}^\dag)$. Let $\mathcal{G}^\dag$ be quasi-strongly connected  and assume that there exists a constant $0<q_\ast<1$ such that $ \mathbf{P} \big( (i,j)\in\mathrm
{E}_k \big) \geq q_\ast$ for all $k$ and $(i,j)\in \mathcal{E}^\dag$. Then Algorithm (\ref{9}) achieves  global a.s. consensus  if and only if $\sum_{k=0}^\infty {P}_k =\infty$, and there holds
\begin{align*}
\mathscr{T}_{\rm com}(\epsilon)\leq \inf\Big\{k:\sum_{i=0}^{k-1}\big(1-(1-P_i)^n\big) \geq \frac{(n-1)|\mathcal{E}^\dag|}{\log A} \log \big(A\epsilon^2/n\big) \Big\}
\end{align*}
where $A=1- \big({ \eta q_\ast}/{n}\big)^{(n-1) |\mathcal{E}^\dag|}$ and $|\mathcal{E}^\dag|$ represents the number of elements in $\mathcal{E}^\dag$.
\end{theorem}

\subsection{Remarks}
Theorems \ref{theorem-Impossibility} and \ref{theorem-ARC-independent} combined show that $\sum_{k=0}^\infty {P}_k =\infty$ is a sharp threshold  for  Algorithm (\ref{9}) to reach  consensus convergence under quite general conditions leading to a {\it Zero-One Law}: consensus convergence holds true with probability one
for all initial conditions when the infinite sum diverges, and otherwise consensus is achieved  with probability zero for almost all initial conditions. These results are essentially due to the Borel-Cantelli Lemma, where the condition $\sum_{k=0}^\infty {P}_k =\infty$ ensures  that a.s. every node updates its states infinitely often.
Consensus convergence however requires much more than that: it is crucial that every node updates  infinitely often  (cf., Theorem \ref{theorem-Impossibility}), but more importantly,  these node updates should essentially comply with the underlying  random graph process (cf., Theorems \ref{theorem-connectivity-independent} and \ref{theorem-ARC-independent}).

The convergence time estimates given in  Theorems \ref{theorem-connectivity-independent} and \ref{theorem-ARC-independent} are established with the help of  Markov's inequality, where the random graph process and the Bernoulli trials together cumulatively contribute to the upper bound.  Moreover, from the proofs it is straightforward  to see that all the results can be easily generalized to the case when nodes have distinct successful update probabilities, i.e., the $P_k$ become $P_k^i$ depending on $i\in \mathrm{V}$. For the ease of presentation we let $\langle P_k \rangle$ apply to all nodes.

The proposed concepts of connectivity and arc independence directly apply   to many random graph models in the literature, for which a detailed discussion is provided
in Subection \ref{Subsec:Examples} illustrating the usefulness of the derived results for several fundamental random graph processes. In fact, we manage to generalize
the  connectivity/arc independence to connectivity/arc {\it mixing} of random graph processes, and show that the same analysis smoothly carries on leading to similar convergence results. As a result, the majority of the random graph models in the literature (including Markovian random graphs) are then covered. The generalization to mixing of random graphs is
in Subsection \ref{Subsec:Mixing}.

Finally, for random gossiping algorithms \cite{boyd}, we are able to go one step forward and present almost sure finite-time convergence results. Remarkably enough, the random node update process driven by  non-reliable  $\langle P_k \rangle$ becomes  {\it essential} for  almost sure finite-time convergence. This part of the generalization is in Subsection \ref{Subsec:Gossiping}.

\subsection{Paper Organization}
The remainder  of the paper is organized as follows. Section \ref{Sec: Connectivity} establishes the convergence results  for connectivity-independent graphs. Several classes of connectivity-independent graphs, e.g., directed, bidirectional and acyclic  graphs,  are investigated, respectively.  The proof of Theorem \ref{theorem-connectivity-independent} is therefore  obtained as a  direct consequence. Section \ref{Sec:ARC} turns to arc-independent graphs and proves Theorem \ref{theorem-ARC-independent}  using a matrix argument. Section \ref{Sec:Generalization}  is devoted to some nontrivial generalizations of  the main results: A few basic random graph models  in the literature are  discussed illustrating the applicability of the derived results; the connectivity/arc $*$-mixing random graphs are introduced with the corresponding convergence results established; conditions for almost sure finite-time convergence of random  gossiping algorithms are presented. Finally some concluding remarks are given in Section \ref{Sec:Conclusion}.

 The proof of Theorem  \ref{theorem-Impossibility} on the consensus  impossibilities is a bit tangential to the a.s. convergence results,  and is therefore  put in Appendix~A.

\subsection*{Notation and Terminologies}
A  (simple) directed graph  $\mathsf
{G}=(\mathsf {V}, \mathsf {E})$, or in short, a digraph,  consists of a finite set
$\mathsf{V}=\{1,\dots,N\}$ of nodes and an arc set
$\mathsf {E}$, where  an element $e=(i,j)\in\mathsf {E}$ denotes   an
{\it arc} from node $i\in \mathsf{V}$  to $j\in\mathsf{V}$ with $i\neq j$.  A directed path between two vertices $v_1$ and $v_k$ in $\mathsf{G}$ is a sequence of distinct nodes
$
v_1v_2\dots v_{k}$
such that for any $m=1,\dots,k-1$, there is an arc from $v_m$ to $v_{m+1}$; $
v_1v_2\dots v_{k}$ is called a semi-path if for any $m=1,\dots,k-1$, either $(v_m, v_{m+1})\in\mathsf{E}$ or $(v_{m+1}, v_{m})\in\mathsf{E}$. As is usual \cite{god}, a graph $\mathsf{G}$ is termed  {\it strongly  connected} if, for every pair of distinct nodes in $\mathsf{V}$, there is a path  from one to the other;  {\it quasi-strongly connected} if there exists a node $v\in\mathcal{V}$,  namely  a {\it root},  such that there is a path from $v$ to all other nodes (equivalently, the graph contains a directed spanning tree); {\it weakly connected} if there is a semi-path between any two distinct nodes. Deterministic graphs are denoted by $\mathcal{G}$ with arc set $\mathcal {E}$; random graphs are denoted by $\mathrm{G}$ with arc set $\mathrm{E}$.

We use $\mathbb{N}$ to denote the set of non-negative integers, and $\mathbb{R}$ denotes real numbers. Probability is denoted as $\mathbf{P}(\cdot)$; expectation of random variables is denoted as $\mathbf{E}[\cdot]$; events are denoted by $\mathpzc{A}, \mathpzc{B}, \dots$. We also use $|\cdot|$ to represent the cardinality of a finite set, or the absolute value of a real number.  A sequence $\{b_k\}_{k=0}^{\infty}$  of real numbers, random variables, or  events,  is always abbreviated as $\langle b_k \rangle $.

\section{Connectivity-Independent Graphs}\label{Sec: Connectivity}
In this section, we present the convergence analysis for  connectivity-independent random  graph processes. We are going to study some  general cases  relying on the joint graphs only.

The joint graph \cite{tsi,jad03} of $\mathrm
{G}_k$ on
time interval $[k_1,k_2]$ for $0\leq k_1\leq k_2\leq \infty$ is defined as  $
\mathrm {G}\big([k_1,k_2]\big)=\big(\mathrm {V},\mcup_{k\in[k_1,k_2]}\mathrm
{E}_k\big).$
We introduce the following  definition of connectivity for the random graph process.
 \begin{definition} The random graph process $\langle \mathrm{G}_k\rangle$ is termed

\noindent (i)  {\it uniformly stochastically quasi-strongly connected},
if there exist an integer  $B\geq1$ and $0<q<1$ such that the sequence  $\big\langle \mathrm {G}\big([mB,(m+1)B-1]\big)\big\rangle$ is connectivity-independent and
 $$
 \mathbf{P} \Big( \mathrm {G}\big([mB,(m+1)B-1] \big)\mbox{ is quasi-strongly connected}\Big) \geq q,\ \ \ m=0,1,\dots;
 $$

\noindent (ii)  {\it infinitely stochastically   quasi-strongly connected},  if  there exist a (deterministic) sequence $
0= C_0<\dots<C_m<\dots $ and a constant $0<q<1$ such that  $\Big\langle \mathrm {G}\big([C_m,C_{m+1})\big)\Big \rangle$ is connectivity-independent and
$$
\mathbf{P} \Big(\mathrm {G}\big([C_m,C_{m+1})\big)\mbox{ is quasi-strongly connected}\Big) \geq q,\ \ \ m=0,1,\dots.
$$
\end{definition}

In the remainder of this section, we first establish consensus conditions for  uniformly  stochastically  quasi-strongly connected graph processes, and the proof of Theorem \ref{theorem-connectivity-independent} is obtained as a special case. Next, two special cases,   bidirectional and acyclic graph processes, are further  investigated, respectively.

\subsection{Uniform  Stochastic Connectivity}
The following result is for consensus convergence over uniformly stochastically  quasi-strongly connected graphs.
\begin{proposition}\label{prop1}
Suppose  $\langle \mathrm{G}_k\rangle$ is  uniformly stochastically quasi-strongly connected with  $B\geq1$ and $q>0$. Then Algorithm (\ref{9}) achieves  global a.s. consensus  if $\sum_{s=0}^\infty \bar{P}_s =\infty$, where
$$
\bar{P}_s=\inf_{\alpha_1<\dots<\alpha_{n-1}} \Big\{\prod_{l=1}^{n-1}P_{\alpha_l}:\ \alpha_l\in\big[s(n-1)^2B, (s+1)(n-1)^2B\big), l=1,\dots,n-1\Big\}.
$$
Moreover, we have
$$
\mathscr{T}_{\rm com}(\epsilon) \leq\inf\Big\{M:\ \  \sum_{i=0}^{M-1}\log \Big(1-\frac{(q\eta/n)^{(n-1)^2}}{2}\cdot \bar{P}_i\Big)^{-1}\geq {\log \epsilon^{-2}}\Big\}\times (n-1)^2B,
$$
where $\eta$ is the constant defined in the weights rule.
\end{proposition}
{\it Proof.} We first establish a lemma characterizing a useful property of  uniformly stochastically  quasi-strongly connected graphs.
\begin{lemma}\label{lem3}Assume that $\langle \mathrm{G}_k\rangle$ is uniformly stochastically  quasi-strongly connected. Then for any $s=0,1,\dots$, we have
 \begin{align}
 &\mathbf{P} \Big(\exists i_0\in\mathrm{V}\ {\rm and}\ \tau_1<\dots<\tau_{n-1}\in[s(n-1)^2,(s+1)(n-1)^2)\  \nonumber\\
  &\ \ \ \ \ \ \ \ \ \ \ \ \ \ \ \ \ \ \ {\rm s.t.}\  \mbox{\rm $i_0$ is a root of $\mathrm {G}\big([\tau_j B, (\tau_j+1)B-1]\big)\ {\rm for\ all}\ j=1,\dots,n-1$}\Big)\geq \Big(\frac{q}{n}\Big)^{(n-1)^2}. \nonumber
 \end{align}
\end{lemma}
{\it Proof.} The probability that the  graph $\mathrm {G}\big([\tau B, (\tau+1)B-1]\big)$ has a root is no less than $q$ for $\tau=s(n-1)^2, \dots, (s+1)(n-1)^2-1$. Consequently, for each $\mathrm {G}\big([\tau B, (\tau+1)B-1]\big)$, there is a (deterministic) node $i_0$ (which depends on $\tau$) such that $i_0$ is a root with probability at least $q/n$.  The $(n-1)^2$ graphs, $\mathrm {G}\big([\tau B, (\tau+1)B-1]\big)$, $\tau=s(n-1)^2, \dots, (s+1)(n-1)^2-1$ will lead to at least $(n-1)^2$ such roots  (with possible repetitions, of course). However, the total number of the nodes is $n$. Thus, at least one node is counted more than $n-2$ times. Then the desired lemma immediately follows from the definition of connectivity independence. \hfill$\square$.

 Denote $k_s=s(n-1)^2B$ for $s\geq0$. We consider the event
studied in Lemma \ref{lem3}, that there exists a common root node $i_0$ such that  $i_0$ is a root of $\mathrm {G}\big([\tau_jB, (\tau_j+1)B-1]\big)$ for $ j=1,\dots,n-1$ with $k_s\leq \tau_jB\leq k_{s+1}-1$ with a probability no smaller than $  \big({q}/{n}\big)^{(n-1)^2}$.  We first assume that
\begin{align}\label{a1}
x_{i_0}(k_s)\leq \frac{1}{2} h(k_s)+\frac{1}{2} H(k_s).
\end{align}
From a symmetric analysis by establishing the lower bound of $h(k_{s+1})$ (or, directly  considering $y(k)=(y_1(k) \dots y_n(k))\T$ with $y_i(k)=-x_i(k)$), it is easy to  show  that the argument  continues to hold under the condition  that
\begin{align*}
x_{i_0}(k_s)>\frac{1}{2} h(k_s)+\frac{1}{2} H(k_s).
\end{align*}

We divide the rest of the proof into four steps.

\noindent Step 1.
In this step, we bound $x_{i_0}(k)$. With the weights rule in our standing assumption,  we have
\begin{align}\label{r200}
\sum_{j\in \mathrm{N}_{i_0}(k_s)}a_{i_0j}(k_s)x_j(k_s)
&=a_{i_0i_0}(k_s)x_{i_0}(k_s)+\sum_{j\in \mathrm{N}_{i_0}(k_s)\setminus \{i_0\}}a_{i_0j}(k_s)x_j(k_s)\nonumber\\
&\leq a_{i_0i_0}(k_s)\big(\frac{1}{2}h(k_s)+\frac{1}{2}H(k_s)\big)+ \big(1-a_{i_0i_0}(k_s)\big)H(k_s)\nonumber\\
&= \frac{a_{i_0i_0}(k_s)}{2}h(k_s)+\big(1-\frac{a_{i_0i_0}(k_s)}{2}\big)H(k_s)\nonumber\\
 &\leq\frac{\eta}{2} h(k_s)+\big(1-\frac{\eta}{2}\big)H(k_s),
\end{align}
where the last equality holds from the facts that $ h(k_s)\leq  H(k_s)$ and $a_{i_0i_0}(k_s)\geq \eta$.
Thus, since $\eta<1$, no matter whether  node $i_0$ takes averaging or sticks to its current state, we obtain
\begin{equation*}
x_{i_0}(k_s+1)\leq \frac{\eta}{2} h(k_s)+(1-\frac{\eta}{2})H(k_s).
\end{equation*}
By  recursively applying (\ref{r200}) there holds  that  for any $\varrho=0,1,\dots$,
\begin{equation}\label{1}
x_{i_0}(k_s+\varrho)\leq \frac{\eta^\varrho}{2} h(k_s)+(1-\frac{\eta^\varrho}{2} )H(k_s).
\end{equation}

\noindent Step 2.  Conditioned that $i_0$ is a root of $\mathrm {G}\big([\tau_1B, (\tau_1+1)B-1]\big)$, there will be  $i_1\in\mathrm{V}$, which is different from $i_0$,
 and a time instant $t_1\in[\tau_1B, (\tau_1+1)B-1]$,  such that $(i_0,i_1)\in\mathrm {E}_{t_1}$\footnote{It should be emphasized that
 here both $i_1$ and $t_1$ could be random. We can however treat the remaining analysis for each sample path $\omega$ of
  $i_1(\omega)$ and $t_1(\omega)$, and the same conclusion will be drawn. Therefore, without loss of generality, we omit the discussion regarding
  possible randomness in  $i_1$ and $t_1$.}. Suppose $t_1=k_s+\varsigma$. In this step, we bound $x_{i_1}(k)$.

If $i_1$ takes an averaging update at  time step $t_1$,  we obtain from (\ref{1}) that
\begin{align}
x_{i_1}(k_s+\varsigma +1 )&=a_{i_1i_0}(k_s+\varsigma)x_{i_0}(k_s+\varsigma)+\sum_{j\in \mathrm{N}_{i_1}(k_s+\varsigma)\setminus \{i_0\}}a_{i_1j}(k_s+\varsigma)x_j(k_s+\varsigma)\nonumber\\
 &\leq  \eta \big[\frac{\eta^\varsigma }{2} h(k_s)+(1-\frac{\eta^\varsigma }{2} )H(k_s)\big]+(1-\eta )H(k_s)\nonumber\\
 &=\frac{\eta^{\varsigma+1} }{2} h(k_s)+\big(1-\frac{\eta^{\varsigma+1}}{2} \big)H(k_s),
\end{align}
which leads to that for any $\varrho=(\tau_1+1)B-k_s,\dots$,
\begin{align}\label{r3}
x_{i_1}(k_s+\varrho )
 &\leq
 \frac{\eta^{\varrho} }{2} h(k_s)+(1-\frac{\eta^{\varrho}}{2} )H(k_s).
\end{align}

Therefore, we conclude from (\ref{1}) and (\ref{r3}) that
\begin{equation*}
\mathbf{P} \Big(x_{i_l}(k_s+\varrho )\leq
 \frac{\eta^{\varrho} }{2} h(k_s)+(1-\frac{\eta^{\varrho}}{2} )H(k_s),\; l=0,1;\varrho\geq(\tau_1+1)B-k_s\Big)\geq   \Big(\frac{q}{n}\Big)^{(n-1)^2}\min_{k\in [\tau_1B, (\tau_1+1)B-1] } P_{k}. \nonumber
\end{equation*}

\noindent Step 3.
We proceed the analysis for the time interval $[\tau_2B, (\tau_2+1)B-1]$. Similarly, $i_2\neq i_0,i_1$  can be found (because $i_0$ is a root) such that
\begin{align*}
\mathbf{P} \Big(x_{i_l}(k_s+\varrho )\leq
 \frac{\eta^{\varrho} }{2} h(k_s)+(1-\frac{\eta^{\varrho}}{2} )H(k_s),\;l=0,1,2;\varrho\geq(\tau_2+1)B-k_s\Big) \nonumber\\
 \geq   \Big(\frac{q}{n}\Big)^{(n-1)^2} \min_{\alpha_1,\alpha_2\in [\tau_1B, (\tau_2+1)B-1] } P_{\alpha_1}P_{\alpha_2}.\nonumber
\end{align*}
Continuing for  time intervals $[\tau_jB, (\tau_j+1)B-1]$ for $j=3,\dots, n-1$,  bounds for $i_3, \dots, i_{n-1}$ can be established as
\begin{equation*}
\mathbf{P} \Big(x_{i_m}(k_s+\varrho )\leq
 \frac{\eta^{\varrho} }{2} h(k_s)+(1-\frac{\eta^{\varrho}}{2} )H(k_s),\;m=0,\dots,n-1;\varrho=(\tau_{n-1}+1)B-k_s,\dots\Big) \geq \bar{P}_s   \Big(\frac{q}{n}\Big)^{(n-1)^2}, \nonumber
\end{equation*}
with
$$
\bar{P}_s=\inf_{\alpha_1<\dots<\alpha_{n-1}} \Big\{\prod_{m=1}^{n-1}P_{\alpha_m}:\ \alpha_m\in\big[s(n-1)^2B, (s+1)(n-1)^2B\big), m=1,\dots,n-1\Big\}.
$$
This leads to
\begin{equation}\label{3}
\mathbf{P} \Big(H(k_{s+1} )\leq
 \frac{\eta^{(n-1)^2} }{2} h(k_s)+(1-\frac{\eta^{(n-1)^2}}{2} )H(k_s)\Big) \geq\bar{P}_s  \Big(\frac{q}{n}\Big)^{(n-1)^2},
\end{equation}
which  implies
\begin{equation}\label{s10}
\mathbf{P} \Big(\mathcal{H}(k_{s+1})\leq \Big(1-\frac{\eta^{(n-1)^2}}{2}\Big)\mathcal{H}(k_s)\Big) \geq \bar{P}_s   \Big(\frac{q}{n}\Big)^{(n-1)^2}.
\end{equation}

Therefore, noticing $\mathcal{H}(k+1)\leq \mathcal{H}(k)$ always holds,   we obtain from (\ref{s10}) that
\begin{equation}\label{r18}
\mathbf{E}\big  [\mathcal{H}(k_{s+1})\big ] \leq \Big (1-\frac{(q\eta/n)^{(n-1)^2}}{2}\cdot \bar{P}_s\Big )\mathbf{E} \big [\mathcal{H}(k_s)\big ].
\end{equation}

\noindent Step 4. In light of  connectivity-independence and the fact that (\ref{s10}) holds only requiring the connectivity of the union graph
of $n-1$ subintervals in $[s(n-1)^2,(s+1)(n-1)^2)$, we further conclude from (\ref{r18}) that\footnote{Note that in general recursively applying (\ref{r18}) for $s=0,1,\dots$ requires
independence of the random graph process $\langle \mathrm{G}_k \rangle$ that drives  $x(k)$. However, from the proof of Step 2 it is clear that connectivity independence has been enough for
granting such a recursive analysis and therefore ensuring  (\ref{r2}).}
\begin{equation}\label{r2}
\mathbf{E} \big [ \mathcal{H}(k_{M})\big ] \leq \prod_{s=0}^{M-1}\Big(1-\frac{(q\eta/n)^{(n-1)^2}}{2}\cdot \bar{P}_s\Big)\mathcal{H}(0)
\end{equation}
for all $M\geq1$. Thus, it follows from Lemma \ref{lem1} that
\begin{equation}\label{5}
\lim_{M\rightarrow \infty}\mathbf{E} \big[\mathcal{H}(k_{M})\big] =0,
\end{equation}
which yields $\lim_{k\rightarrow \infty}\mathbf{E} \big[\mathcal{H}(k)\big] =0$
since  $\mathcal{H}(k)$ is non-increasing. Using Fatou's Lemma, we further obtain
\begin{equation}\label{a2}
0\leq\mathbf{E} \big[\lim_{k\rightarrow \infty}\mathcal{H}(k)\big] \leq \lim_{k\rightarrow \infty}\mathbf{E}\big[\mathcal{H}(k)\big] =0.
\end{equation}
Therefore, the convergence claim of the conclusion holds because (\ref{a2}) implies
$$
\mathbf{P}\big (\lim_{k\rightarrow +\infty} \mathcal{H}(k)=0\big)=1.
$$

Applying Markov's Inequality to (\ref{r2}) leads to
\begin{align*}
\mathbf{P}\Big({\mathcal{H}(k_M)}\geq \epsilon {\mathcal{H}(0)}\Big)\leq \frac{1}{\epsilon}\cdot \frac{\mathbf{E}[ \mathcal{H}(k_{M})]}{\mathcal{H}(0)}\leq \frac{1}{\epsilon} \prod_{s=0}^{M-1}\big(1-\frac{(q\eta/n)^{(n-1)^2}}{2}\cdot \bar{P}_s\big).
\end{align*}
Consequently, we have
\begin{align*}
\mathscr{T}_{\rm com}(\epsilon)\leq  \inf\Big\{M: \sum_{s=0}^{M-1}\log \Big(1-\frac{(q\eta/n)^{(n-1)^2}}{2}\cdot \bar{P}_s\Big)^{-1}\geq {\log \epsilon^{-2}}\Big\}\times (n-1)^2B.
\end{align*}
The desired conclusion follows.
\hfill$\square$

\begin{remark}
Suppose $P_{k+1}\leq P_k$ for all $k$. Then it is not hard to see that  $\sum_{s=0}^\infty \bar{P}_s =\infty$ if and only if $\sum_{k=0}^\infty {P}_k^{n-1} =\infty$. Then
 Theorem \ref{theorem-connectivity-independent} holds immediately from Proposition  \ref{prop1}.
\end{remark}
\subsection{Bidirectional Connections}
A digraph $\mathcal {G}=(\mathrm{V},\mathcal{E})$ is  {\em bidirectional}
if for any two nodes $i$ and $j$, $(i,j)\in\mathcal{E}$  if and only if $(j,i)\in\mathcal{E}$. In this subsection, we study   bidirectional graphs.

Note that we do not impose an upper bound for the length of the intervals $[C_m,C_{m+1})$ in the definition of infinitely stochastically  quasi-strong connectivity, which  makes an essential difference from the bounded joint connections.
\begin{proposition}\label{thm2}
Suppose there holds a.s. that $\mathrm
{G}_k\ \mbox{is  bidirectional}$ for all $k\geq0$. Assume that $\langle \mathrm{G}_k\rangle$  is  stochastically  infinitely quasi-strongly connected. Then Algorithm (\ref{9}) achieves global a.s.  consensus if $\sum_{s=0}^\infty \hat{P}_{s}=\infty$, where
$$
\hat{P}_s=\inf_{\alpha_1<\dots<\alpha_{n-1}} \Big\{\prod_{l=1}^{n-1}P_{\alpha_l}:\ \  \alpha_l\in\big[C_{s(n-1)},C_{(s+1)(n-1)}\big), l=1,\dots,n-1\Big\}.
$$
and also
$$
\mathscr{T}_{\rm com}(\epsilon)\leq \inf\Big\{C_{s(n-1)}:\ \  \sum_{i=0}^{s-1}\log \Big(1-(q\eta)^{(n-1)}\cdot \hat{P}_i\Big)^{-1}\geq {\log \epsilon^{-2}}\Big\}.
$$
\end{proposition}
\noindent{\it Proof.} The proof is carried out analyzing a sequence a stopping times (cf., \cite{Durr}) of the random graph process. Take a node $i_0\in\mathrm{V}$ with $x_{i_0}(C_0)=h(C_0)$. Define $$
t_1=\inf_{k\geq C_0}\Big\{\mathrm{N}_{i_0}(k)\setminus\{i_0\}\neq \emptyset\Big\}.
$$
In other words, $t_1$ is the first time that $i_0$ has a neighbor other than itself in the random graph process.  Then $t_1$ is a stopping time of the graph process $\langle \mathrm{G}_k\rangle$, and  there must hold $$
\mathbf{P}\big(t_1<C_1\big) >q
$$
since $\langle \mathrm{G}_k\rangle$ is infinitely  stochastically  quasi-strongly connected. We next introduce
 $$
 \mathrm{V}_1=\Big\{j\neq i_0:\  (i_0,j)\in \mathrm{E}_{t_1}\Big\}.
 $$
We emphasize that by its definition $ \mathrm{V}_1$ is  nonempty. Again we proceed in steps for the remaining of the argument.

\noindent {Step 1.} Noticing that $x_{i_0}(t_1)=x_{i_0}(C_0)$ from the structure of Algorithm (\ref{1}), we conclude that
\begin{align}\label{r11}
x_{i_0}(t_1+1)&=\sum_{j\in \mathrm{N}_{i_0}(t_1)}a_{i_0j}(t_1)x_j(t_1)\nonumber\\
&=a_{i_0i_0}(t_1)x_{i_0}(t_1)+\sum_{j\in \mathrm{N}_{i_0}(t_1)\setminus \{i_0\}}a_{i_0j}(t_1)x_j(t_1)\nonumber\\
&\leq a_{i_0i_0}(t_1)h(C_0)+ (1-a_{i_0i_0}(t_1))H(C_0)\nonumber\\
 &\leq \eta h(C_0)+(1-\eta)H(C_0)
\end{align}
if $i_0$ updates its state at time $t_1+1$. Note that (\ref{r11}) continues to hold even if  $i_0$ fails to update at time $t_1+1$.
On the other hand, for $i_1\in \mathrm{V}_1$,  if  $i_q$ successfully updates its state at time $t_1+1$, then
\begin{align*}
x_{i_1}(t_1+1)&=\sum_{j\in \mathrm{N}_{i_1}(t_1)}a_{i_1j}(t_1)x_j(t_1)\nonumber\\
&=a_{i_1i_0}(t_1)x_{i_0}(t_1)+\sum_{j\in \mathrm{N}_{i_1}(t_1)\setminus \{i_0\}}a_{i_1j}(t_1)x_j(t_1)\nonumber\\
&\leq a_{i_1i_0}(t_1)h(C_0)+ (1-a_{i_1i_0}(t_1))H(C_0)\nonumber\\
 &\leq \eta h(C_0)+(1-\eta)H(C_0),
\end{align*}
which gives
\begin{equation}\label{r5}
\mathbf{P} \Big(x_{i_1}(t_1+1)\leq \eta h(C_0)+(1-\eta)H(C_0)\Big) \geq P_{t_1}.
\end{equation}

The fact that the random node updates are independent of the graph process as well as  among different nodes,  allows us to further conclude that
\begin{equation*}
\mathbf{P} \Big(x_m(t_1+1)\leq \eta h(C_0)+(1-\eta)H(C_0),\ m\in \{i_0\}\mcup\mathrm{V}_1\Big) \geq P_{t_1}^{|\mathrm{V}_1|}.
\end{equation*}

\noindent{Step 2.} We continue to  define $$
t_2=\inf_{k\geq t_1}\Big\{\mbox{there is an arc between $\mathrm{V}\setminus (\{i_0\}\mcup\mathrm{V}_1)$ and }\{i_0\}\mcup\mathrm{V}_1 \mbox{in}\  \mathrm{G}_k\Big\}
$$ and $$
\mathrm{V}_2=\{j\in\mathrm{V}\setminus (\{i_0\}\mcup\mathrm{V}_1):\  \mbox{there is an arc between} \{i_0\}\mcup\mathrm{V}_1\ \mbox{and $j$ in}\ \mathrm{G}_{t_2}\Big\}.
 $$
 Again $t_2$ is a stopping time of the graph process $\langle \mathrm{G}_k\rangle$, and by definition there holds $$
\mathbf{P}\big(t_2<C_2|t_1<C_1\big) \geq q.
$$
Repeating the similar analysis as Step 1 we obtain
\begin{equation*}
\mathbf{P} \Big(x_m(t_2+1)\leq \eta^2 h(C_0)+(1-\eta^2)H(C_0), \ m\in \{i_0\}\mcup\mathrm{V}_1\mcup\mathrm{V}_2 \Big) \geq P_{t_1}^{|\mathrm{V}_1|} P_{t_2}^{|\mathrm{V}_2|}.
\end{equation*}

\noindent {Step 3.} Continuing the analysis, $t_3,\dots,t_{\mu_0}$ and $\mathrm{V}_3,\dots\mathrm{V}_{\mu_0}$ can be found until $$
\mathrm{V}=\{i_0\}\mcup\big(\mcup_{l=1}^{\mu_0}\mathrm{V}_l\big)
$$ for some $\mu_0\leq n-1$, and eventually
\begin{equation*}
\mathbf{P} \Big(x_m(t_{\mu_0}+1)\leq \eta^{\mu_0} h(C_0)+(1-\eta^{\mu_0})H(C_0), \ m\in \mathrm{V} \Big) \geq \prod_{l=1}^{\mu_0} P_{t_l}^{|\mathrm{V}_l|}.
\end{equation*}
This gives us
\begin{equation}\label{r12}
\mathbf{P} \Big(\mathcal{H}(t_{\mu_0}+1)\leq \Big(1-\eta^{\mu_0}\Big) \mathcal{H}(C_0) \Big) \geq \prod_{l=1}^{\mu_0} P_{t_l}^{|\mathrm{V}_l|}.
\end{equation}
Noting that $\mu_0\leq n-1$ as well as the fact that $\mathbf{P} \big(t_{\mu_0}<C_{n-1}\big) \geq q^{n-1}$ in light of the connectivity independence, we further conclude from (\ref{r12}) that
\begin{equation*}
\mathbf{P} \Big(\mathcal{H}(C_{n-1})\leq (1-\eta^{n-1}) \mathcal{H}(C_0) \Big) \geq  \hat{P}_0 q^{n-1}
\end{equation*}
with $
\hat{P}_0=\inf_{\alpha_1<\dots<\alpha_{n-1}} \Big\{\prod_{l=1}^{n-1}P_{\alpha_l}:\ \  \alpha_j\in\big[C_{0},C_{n-1}\big), j=1,\dots,n-1\Big\}.$

We can recursively apply the above analysis and bounds of  $\mathcal{H}(C_{s(n-1)})$ can be  obtained for $s=1,2,\dots$ as
\begin{equation*}
\mathbf{P} \Big(\mathcal{H}\big(C_{(s+1)(n-1)}\big)\leq (1-\eta^{n-1}) \mathcal{H}\big(C_{s(n-1)}\big) \Big) \geq  \hat{P}_s q^{n-1}.
\end{equation*}
This ensures global a.s. consensus  and that
\begin{align*}
\mathscr{T}_{\rm com}(\epsilon)\leq  \inf\Big\{C_{s(n-1)}:\ \  \sum_{i=0}^{s-1}\log \Big(1-(q\eta)^{(n-1)}\cdot \hat{P}_i\Big)^{-1}\geq {\log \epsilon^{-2}}\Big\}.
\end{align*}
 The proof is now complete. \hfill$\square$

\begin{remark}
If we impose the condition that $P_{k+1}\leq P_k$ for all $k$ in  Proposition \ref{thm2}, then  global a.s. consensus for Algorithm (\ref{9}) can be guaranteed by  $\sum_{s=0}^\infty {P}_{_{C_{s(n-1)}}}^{n-1}=\infty$.
\end{remark}

\subsection{Acyclic Connections}
A digraph $\mathcal{G}=(\mathrm{V}, \mathcal{E})$ is  called {\it acyclic} if it contains no  directed cycle. This subsection focuses on acyclic graphs.

Let $\mathcal{G}$ be an acyclic, quasi-strongly connected digraph. Then it is not hard to see that $\mathcal{G}$ has one and only one root.
Denote this root as $v_0$ and let $v_0\rightarrow j$ be a path from $v_0$ to $j$ in $\mathcal{G}$. Let $|v_0\rightarrow j|$ represent the length of $v_0\rightarrow j$
as the number of arcs in this path.
 We now define a function on the vertices of $\mathcal{G}$ by $\hbar(v_0)=0$ and $\hbar(j)=\max\{|v_0\rightarrow j|:$ $v_0\rightarrow j$ is a path in $\mathcal{G}\}$ for any $j\neq v_0$.
  Let $d_\ast=\max_{i\in \mathrm{V}}\hbar(i)$, which is the maximum path length and therefore is upper bounded by $n-1$. Then we establish the following lemma indicating that this function $\hbar$ is surjective from $\mathrm{V}$ to $\{0,\dots,d_\ast\}$.

\begin{lemma}\label{lem4}
Let $\mathcal{G}$ be an acyclic, quasi-strongly connected digraph. Then  $\hbar^{-1}(m)=\{i:\ \hbar(i)=m\}$ is nonempty for all $m=0,\dots,d_\ast$.
\end{lemma}
\noindent {\it Proof.} The conclusion holds for $m=0$ trivially.

Let us prove the conclusion  for $m=1$ by contradiction. Assume that  $\hbar^{-1}(1)=\emptyset$. Then we have $m_0\doteq\inf_{i\neq v_0}\hbar(i)>1$.
Take a node $j_0$ with $\hbar(j_0)=m_0$.  There exists a (simple) path $v_0\rightarrow j_0$ in $\mathcal{G}$ with length  $m_0>1$. Let $v_\ast$
be the node for which arc $(v_\ast,j_0)$ is included in $v_0\rightarrow j_0$.   According to the definition of $m_0$, we have $\hbar(v_\ast)\geq m_0$.
Suppose $v_0\rightarrow v_\ast$ is a path with length $\hbar(v_\ast)$. Note that, $j_0$ cannot be included in $v_0\rightarrow v_\ast$ because then it
 would generate a cycle $j_0\rightarrow v_\ast \rightarrow j_0$. Consequently,  another path $v_0\rightarrow v_\ast\rightarrow j_0$ is obtained whose length
 is $\hbar(v_\ast)+1>m_0$. This contradicts the selection rule of $j_0$. Therefore, the conclusion holds for $m=1$.

Next, we construct another graph ${\mathcal{G}}^\ast$ from $\mathcal{G}$ by viewing node set $\{v_0\}\mcup\hbar^{-1}(1)$ as a single node
in the new graph without changing the arcs. We see that $\hbar^{-1}(2)$ of $\mathcal{G}$ is exactly the same as the node set $\hbar^{-1}(1)$ of ${\mathcal{G}}^\ast$, while the latter is nonempty via previous analysis. Continuing the argument, the conclusion follows. \hfill$\square$

 Here comes our main result for acyclic graphs.
\begin{proposition}\label{thm3}
Let $\mathcal{G}=(\mathrm{V},\mathcal{E})$ be an acyclic, quasi-strongly connected digraph and assume that a.s. $\mathrm{E}_k  \subseteq \mathcal{E}$ for all $k$.
Suppose  $\langle \mathrm{G}_k\rangle$ is  infinitely   stochastically quasi-strongly connected. Then Algorithm (\ref{9}) achieves  global  a.s. consensus if $\sum_{s=0}^\infty \big[ \inf_{C_s\leq\alpha<C_{s+1}}P_{\alpha} \big]=\infty$.
\end{proposition}
{\noindent \it Proof.} Let $v_0$ be the unique root node of $\mathcal{G}$.  Based on Lemma \ref{lem4}, $  {\mathrm{V}}_i=\hbar^{-1}(i)$ for $i=0,\dots,\hbar_0$ can be defined satisfying   $  {\mathrm{V}}_0=\{v_0\}$ and ${\mathrm{V}}=\mcup_{i=0}^{d_\ast}   {\mathrm{V}}_i$. Obviously we have $\mathbf{P} \big(x_{v_0}(k)=x_{v_0}(0), k\geq 0\big) =1$  because with probability one, $v_0$ has no neighbor except itself for all $k$. We first  investigate the nodes in  $  {\mathrm{V}}_1$.

\vspace{2mm}

\noindent {\it Claim.} $\mathbf{P} \Big(\lim_{k\rightarrow \infty}|x_{m}(k)-x_{v_0}(0)|=0 \Big) =1$  for all $m\in  {\mathrm{V}}_1$.

Take $v_1\in  {\mathrm{V}}_1$. By our assumption $v_0$ is the only neighbor of $v_1$ excluding itself in $\mathrm{G}_k$ for all $k$. Define $t_1=\inf_{k\geq0}\{(v_0,v_1)\in\mathrm{E}_k\}$. Then $\mathbf{P} (t_1<C_1) \geq q$. We have
\begin{align*}
  \Big|\sum_{j\in \mathrm{N}_{v_1}(t_1)}a_{v_1j}(t_1) x_j(t_1)-x_{v_0}(0)\Big|&=\big|a_{v_1v_0}(t_1)x_{v_0}(t_1)+ a_{v_1v_1}(t_1)x_{v_1}(t_1)-x_{v_0}(0)\big|\nonumber\\&=\big(1-a_{v_1v_0}(t_1)\big)\big|x_{v_1}(0)-x_{v_0}(0)\big|\nonumber\\
  &\leq (1-\eta)\big|x_{v_1}(0)-x_{v_0}(0)\big|,
\end{align*}
which yields
\begin{equation*}
\mathbf{P} \Big(\big|x_{v_1}(t_1+1)-x_{v_0}(0)\big|\leq(1-\eta)\big|x_{v_1}(0)-x_{v_0}(0)\big|\Big) \geq P_{t_1}.
\end{equation*}
Thus, we obtain
\begin{equation*}
\mathbf{P} \Big(\big|x_{v_1}(C_1)-x_{v_0}(0)\big|\leq(1-\eta)\big|x_{v_1}(0)-x_{v_0}(0)\big|\Big) \geq \tilde{P}_{0}q,
\end{equation*}
where $\tilde{P}_{0}=\inf_{C_0\leq\alpha<C_{1}}P_{\alpha}$. Repeating the analysis on time interval $[C_{m},C_{m+1})$, $m=1,2,\dots$, we have
\begin{equation}\label{16}
\mathbf{P} \Big(\big|x_{v_1}(C_{m+1})-x_{v_0}(0)\big|\leq(1-\eta)\big|x_{v_1}(C_m)-x_{v_0}(0)\big|\Big) \geq \tilde{P}_{m}q,\ \  m=1,2,\dots.
\end{equation}
Then connectivity independence  leads to $\mathbf{P} \big(\lim_{k\rightarrow \infty}|x_{v_1}(k)-x_{v_0}(0)|=0 \big) =1$. The claim is proved.

\vspace{2mm}

Next, we turn to $  {\mathrm{V}}_2$. The analysis of nodes in $\mathrm{V}_2$ will be carried out via a sample-path argument. Let $\ell\geq 1$. The above  claim we just proved implies that there exists a random variable $T_\ell(\omega)>0$ which is a.s. finite, such that
$$
\mathbf{P} \Big(|x_{m}(k)-x_{v_0}(0)|\leq 1/\ell, \ k\geq T_\ell, m\in  {\mathrm{V}}_1\Big) =1.
 $$
Take $v_2$ as an arbitrary node in $  {\mathrm {V}}_2$. We prove  $\mathbf{P} \big(\limsup_{k\rightarrow \infty}|x_{v_2}(k)-x_{v_0}(0)|\leq 1/\ell \big) =1$. Note that from the definition of $\mathrm{V}_2$, the state of node $v_2$ can only be influenced by nodes in $\{v_0\}\mcup\mathrm{V}_1$ in the entire time horizon.

Define event  $\mathscr{E}=\big\{|x_{v_2}(k)-x_{v_0}(0)|>1/\ell,  \ k\geq T_\ell\big\}$. There will be two cases.
\begin{itemize}
\item[(i)] Since $v_2$ can only possibly connect to  nodes in $\{v_0\}\mcup\mathrm{V}_1$, there must hold $$
|x_{v_2}(s)-x_{v_0}(0)|\leq 1/\ell,\ s\geq k
$$ if  $|x_{v_2}(k)-x_{v_0}(0)|\leq 1/\ell$ for $k\geq T_\ell$.  This is equivalent to that  $$
\mathbf{P} \big(\limsup_{k\rightarrow \infty}|x_{v_2}(k)-x_{v_0}(0)|\leq 1/\ell |\mathscr{E}^c\big) =1.
$$

\item[(ii)] By our assumption,  $\langle \mathrm{G}_k\rangle$ is  infinitely   stochastically quasi-strongly connected, and  $\sum_{s=0}^\infty \tilde{P}_{s}=\infty$ with $\tilde{P}_s=\inf_{C_s\leq\alpha<C_{s+1}}P_{\alpha}$. Moreover, whenever $\mathrm {G}\big([C_m,C_{m+1})\big)$ is quasi-strongly connected, there must be
a time instant between $[C_m,C_{m+1})$ at which  there is an arc from $\{v_0\}\mcup   {\mathrm{V}}_1$ to  $v_2$. Therefore, applying the Borel-Cantelli lemma we see that a.s. there is an infinite sequence of (random) time instants
$$
t_1<t_2<\cdots<t_m< \cdots
$$
where  at each  time $t_m$,  there is at least one arc  from $\{v_0\}\mcup   {\mathrm{V}}_1$ to  $v_2$ and $v_2$ successfully updates its state. Then we obtain for all  $t_m\geq T_\ell$ that
\begin{align*}
& \Big| x_{v_2}(t_m+1)-x_{v_0}(0)\Big|\nonumber\\
&= \Big|\sum_{j\in \mathrm{N}_{v_2}({t}_m)}a_{v_2j}({t}_m) x_j(t_m)-x_{v_0}(0)\Big|\nonumber\\
&= \sum_{j\in\mathrm{N}_{v_2}({t}_m): j\in \{v_0\}\mcup   {\mathrm{V}}_1}a_{v_2j}(t_m) \big|x_j(t_m)-x_{v_0}(0)\big|+a_{v_2v_2}(t_m)\big |x_{v_2}(t_m)-x_{v_0}(0)\big|\nonumber\\
&\leq  \big[1-a_{v_2v_2}(t_m)\big]\frac{1}{\ell}+a_{v_2v_2}(t_m) \big|x_{v_2}(t_m)-x_{v_0}(0)\big|\nonumber\\
&\leq \big[1-\eta\big]\frac{1}{\ell}+\eta \big|x_{v_2}(t_m)-x_{v_0}(0)\big|,
\end{align*}
which reads
\begin{align}\label{32}
\Big|x_{v_2}(t_m+1)-x_{v_0}(0)\Big|-\frac{1}{\ell}
&\leq \eta \Big(\big|x_{v_2}(t_m)-x_{v_0}(0)\big|-\frac{1}{\ell}\Big)
\end{align}
for all  $t_m\geq T_\ell$. Noticing the definition of $t_m$, this has already proved that
\begin{equation*}
\mathbf{P} \Big( \limsup_{k\rightarrow \infty}|x_{v_2}(k)-x_{v_0}(0)|\leq 1/\ell\Big|\mathscr{E}\Big)  =1.
\end{equation*}
\end{itemize}
As a result, we can now conclude that
$$
\mathbf{P} \Big(\limsup_{k\rightarrow \infty}|x_{v_2}(k)-x_{v_0}(0)|\leq 1/\ell \Big) =1.
$$
Now that $\ell$ is chosen arbitrarily and $v_2$ is an arbitrary node in $\mathrm{V}_2$, we further obtain
$$
\mathbf{P} \Big(\lim_{k\rightarrow \infty}|x_{m}(k)-x_{v_0}(0)|=0 \Big) =1, \; \; m\in \mathrm {V}_2.
$$

\medskip

Finally, a recursive analysis for   node sets $  {\mathrm{V}}_3,\dots,  {\mathrm{V}}_{d_\ast}$ eventually gives  $\mathbf{P} \big(\lim_{k\rightarrow \infty}$
$|x_{m}(k)-x_{v_0}(0)|=0 \big) =1$ for all $m\in \mathrm {V}$. The desired conclusion thus follows.  \hfill$\square$

\begin{remark}
Proposition  \ref{thm3} immediately  implies the following interesting observations   when  there holds  $\mathrm{E}_k  \subseteq \mathcal{E}$ for all $k$ for some acyclic, quasi-strongly connected digraph $\mathcal{G}=(\mathrm{V}, \mathcal{E})$.
\begin{itemize}
\item[(i).] If   $\langle \mathrm{G}_k\rangle$ is  infinitely   stochastically quasi-strongly connected,  and $P_{k+1}\leq P_k$ for all $k$,
then Algorithm  (\ref{9}) achieves global  a.s. consensus when $\sum_{m=0}^\infty {P}_{C_{m}}=\infty$.

\item[(ii).] Suppose   $\langle \mathrm{G}_k\rangle$ is   uniformly  stochastically quasi-strongly connected with respect to $B\geq 1$. Let either $B=1$ or  $P_{k+1}\leq P_k$, $k\geq 0$.
  Then Algorithm (\ref{9}) achieves  global  a.s. consensus  if and only if $\sum_{k=0}^\infty {P}_{k}=\infty$.
\end{itemize}
\end{remark}
\section{Arc-Independent Graphs}\label{Sec:ARC}
In this section, we investigate the convergence with an arc-independent random graph process and prove Theorem \ref{theorem-ARC-independent}.  We present the convergence analysis
  using a stochastic matrix argument.

Let $e_i=(0 \dots 1 \dots\ 0)\T$ be the $n\times1$ unit vector with the $i$th component equal to one. Denote $\bar{a}_i(k)=(\bar{a}_{i1}(k)\ \dots\ \bar{a}_{in}(k))\T$  as an  $n\times1$  vector with $\bar{a}_{ij}(k)= a_{ij}(k)$ if $j\in \mathrm{N}_i(k)$, and $\bar{a}_{ij}(k)=0$ otherwise.
Let $\chi_i(k), i=1,\dots,n, k\in\mathbb{N}$ be a sequence of independent Bernoulli random variables satisfying $\mathbf{E}[\chi_i(k)]=P_k$.
Denote  $W(k)=(w_1(k) \dots w_n (k))\T\in \mathbb{R}^{n\times n}$ as a random matrix with
\begin{equation}
	w_i(k) = \chi_i(k)	\bar{a}_i(k)+(1-\chi_i(k))e_i,\ i=1,\dots,n.
\end{equation}
 Algorithm (\ref{9}) can be written  into a compact form:
\begin{equation}\label{matrix}
x(k+1)=W(k)x(k).
\end{equation}

In the remainder  of this section, we first establish several useful lemmas on the product of stochastic matrices, and then the proof  of Theorem \ref{theorem-ARC-independent} is presented.

\subsection{Stochastic Matrices}
A finite square matrix $M=[M_{ij}]\in\mathbb{R}^{n\times n}$ is called {\em stochastic} if $M_{ij}\geq 0$ for all $i,j$ and $\sum_{j=1}^n M_{ij}=1$ for all $i$ \cite{lat}. For a stochastic matrix $M$, introduce
\begin{equation*}
\delta (M)=\max_j \max_{\alpha, \beta}|M_{\alpha j}-M_{\beta j}|, \quad \lambda (M)=1-\min_{\alpha, \beta}\sum_j \min \{M_{\alpha j},M_{\beta j}\}.
\end{equation*}
The $\lambda(M)$ is often called the Dobrushin's ergodicity coefficient \cite{Seneta}. Note that $\delta(M)$ describes  how much the weights assigned by two nodes $\alpha,\beta$ to a node $j$ can differ, and $\lambda(M)=0$ if and only if for each pair of nodes $\alpha$ and $\beta$, for each node $j$ either $\alpha$ or $\beta$ has no weight to $j$. If $\lambda (M)<1$,  $M$ a called a {\em scrambling} matrix. A scrambling matrix does not have  two orthogonal rows, which means
that any two rows share   at least one column,  where both of them have strictly positive elements.   The following lemma  can be found in \cite{haj}.

\begin{lemma}\label{lem0}For any $k$ ($k\geq 1$) stochastic matrices $M_1,\dots,M_k$, there holds
\begin{equation*}
\delta (M_1M_2\dots M_k)\leq \prod_{i=1}^{k}\lambda (M_i).
\end{equation*}
\end{lemma}

Let $M=[M_{ij}]\in\mathbb{R}^{n\times n}$ be a matrix with nonnegative entries. We can associate a unique digraph  $\mathcal{G}_M=(\mathrm{V},\mathcal{E}_M)$
with $M$ on node set $\mathrm{V}=\{1,\dots,n\}$ such that $(j,i)\in\mathcal{E}_M$ if and only if $M_{ij}>0$.
We call $\mathcal{G}_M$ the {\em induced graph} of $M$.

The following lemma is given on the induced graphs  of products of stochastic matrices.

\begin{lemma}\label{lem5}
Let $M_1,\dots,M_k$ be $k\geq1$  stochastic matrices with positive diagonal entries. Then  $\big(\mcup _{i=1}^k \mathrm{G}_{ M_i}\big) \subseteq \mathrm{G}_{ M_k\cdots M_1}$.
\end{lemma}
{\it Proof.} We prove the case  $k=2$, and the conclusion will then follow by induction.

Denote $[{M}_1]_{ij}$, $[{M}_2]_{ij}$ and  $[{M_2 M_1}]_{ij}$ as the $ij$-entries of $M_1$,  $M_2$ and $M_2M_1$, respectively.  Note that, we have
\begin{equation*}
[{M_2 M_1}]_{ij}=\sum_{m=1}^n[{M}_2]_{im}[{M}_1]_{mj}\geq [{M}_2]_{ii}[{M}_1]_{ij} + [{M}_2]_{ij}[{M}_1]_{jj}.
\end{equation*}
The desired conclusion follows immediately from the fact that  $[{M}_2]_{ii}, [{M}_1]_{jj}>0$. \hfill$\square$

The following lemma helps in determining  whether a product of  stochastic matrices is a scrambling matrix.
\begin{lemma}\label{lem6}
Let $M_1, \dots, M_{n-1}$ be $n-1$ stochastic matrices  with positive diagonal entries. Suppose there exists a node $i_0\in\mathrm{V}$ such that $i_0$ is a root  of
$\mathcal{G}_{M_\tau}$ for all $ \tau=1,\dots,n-1$.  Then $M_{n-1}\cdots M_1$ is a  scrambling matrix.
\end{lemma}
{\it Proof.} Since $i_0$ is a root of $\mathcal{G}_{M_1}$, at least one node $i_1$ exists different from  $i_0$ such that $(i_0,i_1)\in \mathcal{E}_{M_1}$. This  immediately implies $[{M}_1]_{i_1i_0}>0$ according to the definition of induced graph.

Further, according to Lemma \ref{lem5}, we have $[M_2M_1]_{i_1i_0}>0$ resulting from $[{M}_1]_{i_1i_0}>0$.  Since $i_0$ is also a root of $\mathcal{G}_{M_2}$, there must be a node $i_2$ different with $i_0$ and $i_1$ such that there is at least one arc leaving from $\{i_0,i_1\}$ entering $i_2$ in $\mathcal{G}_{M_2}$. There will be two cases.
 \begin{itemize}
\item[(i)]  When $(i_0, i_2)\in  \mathcal{E}_{M_2}$, there holds $[M_2]_{i_2i_0}>0$. This implies  $[M_2M_1]_{i_2i_0}>0$ based on Lemma \ref{lem5}.

\item[(ii)] When $(i_1, i_2)\in  \mathcal{E}_{M_2}$, there holds $[M_2]_{i_2i_1}>0$. In this case we have
      \begin{align*}
      [M_2M_1]_{i_2i_0}=\sum_{\tau=1}^n [M_2]_{i_2\tau}  [M_1]_{\tau i_0}\geq [M_2]_{i_2i_1} [M_1]_{i_1i_0} >0.
      \end{align*}
\end{itemize}

We recursively carry out the above analysis, and then  $i_3, \dots, i_{n-1}$ can be found with $\mathrm{V}=\{i_0,\dots,i_{n-1}\}$ such that
\begin{align}\label{19}
[{M}_{n-1}\cdots M_1]_{i_{m} i_0}>0
\end{align}
for all $ m=0,1,\dots,n-1$.
According to the definition of $\delta(\cdot)$, (\ref{19}) implies
$$
\lambda(M_{n-1}\cdots M_1)\leq 1- \min_{m=0,\dots,n-1}[{M}_{n-1}\cdots M_1]_{i_{m} i_0}<1.
$$
The desired lemma  follows. \hfill$\square$


\subsection{Proof of  Theorem \ref{theorem-ARC-independent}: Convergence}
This subsection  presents the proof of the conclusion on a.s. consensus  in  Theorem \ref{theorem-ARC-independent}.  We only need to show the sufficiency part. Note that  global a.s. consensus of (\ref{9}) is  equivalent to
$$
\mathbf{P} \Big(\lim_{k\rightarrow\infty}\delta \big(W(k)\cdots W(0)\big)=0\Big)=1.
$$

We define $	\Psi_k =\mathrm{I}_{\sum_{i=1}^n \chi_k}$ with $\mathrm{I}$ representing the indicator function. In other words,
\begin{equation*}
	\Psi_k =
	\begin{cases}
		1, & \text{if at least one $\chi_i(k)=1$ at $k$;}\\
		0, & \text{otherwise.}\\
	\end{cases}
\end{equation*}
Then we have $\Psi_k=1$ with probability $1-(1-P_k)^n$ and $\Psi_k=0$ with probability $(1-P_k)^n$. Moreover, $\Psi_0,\Psi_1,\dots$ are independent.
Applying the Borel-Cantelli Lemma to $\big\langle \chi_i(k)\big\rangle,i=1,\dots,n$  and invoking the independence $\Psi_k$ we conclude the following property of $\Psi_k$.
\begin{lemma}\label{lem7}
$\mathbf{P} \big(\Psi_k=1$ for infinitely many $k\big) $=1 if and only if $\sum_{k=0}^\infty P_k=\infty$.
\end{lemma}

Noting the fact that
$$
1-ny\leq (1-y)^n
$$
for all $y\in [0,1]$ and $n\geq1$, there holds for all $k$ that
$$
1-(1-P_k)^n \leq nP_k.
$$
As a result, we obtain
\begin{equation}\label{20}
\mathbf{P} \Big(\chi_i(k)=1\Big|\Psi_k=1\Big) =\frac{P_k}{1-(1-P_k)^n}\geq \frac{P_k}{nP_k}=\frac{1}{n}
\end{equation}
for all $i$ and $k$.

Next, we introduce the following sequence (known as the  Bernoulli  sequence of $\langle \Psi_k\rangle$):
\begin{align}\label{a3}
\zeta_1< \ldots <\zeta_m<\zeta_{m+1}<\dots,
\end{align}
where $\zeta_m$ is the $m$'th time when $\Psi_k=1$. The $\zeta_m$ are stopping times of the sequence  $\langle \Psi_k\rangle$, and Lemma \ref{lem6} ensures that each $\zeta_m$ is almost surely finite when $\sum_{k=0}^\infty P_k=\infty$.  For any  $(i,j)\in \mathcal{E}^\dag$, there hold for all  $m=1,2,\dots$ that
\begin{align}\label{36}
&\mathbf{P} \big((i,j)\in \mathrm{G}_{W(\zeta_m)}\big)\nonumber\\
&\stackrel{a)}{=}\mathbf{P} \Big(\chi_{j}(\zeta_m)=1\ \mbox{and}\ (i,j)\in \mathrm{G}_{\zeta_m}\Big)\nonumber\\
&\stackrel{b)}=\mathbf{P} \big(\chi_{j}(\zeta_m)=1\big) \cdot \mathbf{P} \big((i,j)\in \mathrm{G}_{\zeta_m}\big)\nonumber\\
&\stackrel{c)}\geq \frac{q_\ast }{n},
\end{align}
where $a)$ is from the structure  of the algorithm, $b)$ is from the independence between the random graph process and the $\chi_i(k)$, and $c)$ holds from (\ref{20}).

 Denote $Q_1=W(\zeta_{|\mathcal{E}^\dag|})\cdots W(\zeta_{2}) W(\zeta_{1})$, where $|\mathcal{E}^\dag|$ represents the number of elements in $\mathcal{E}^\dag$. From (\ref{36}),
 we can pick up $(i_\tau,j_\tau)$, $\tau=1,\dots,|\mathcal{E}^\dag|$ as all the arcs in $\mathcal{E}^\dag$, and
 the arc-independence (cf., the $\zeta_m$ are stopping times of i.i.d. random process
 $\langle \Psi_k\rangle$, and the random graph process $\langle \mathrm{G}_k\rangle$  is independent with the  $\Psi_k$) leads to
\begin{align*}
\mathbf{P} \Big((i_\tau,j_\tau)\in \mathrm{G}_{W(\zeta_\tau)}, \;\tau=1,\dots,|\mathcal{E}^\dag|\Big) \geq \Big(\frac{q_\ast }{n}\Big)^{|\mathcal{E}^\dag|},
\end{align*}
which yields
\begin{align*}
\mathbf{P} \Big(\mathcal{G}^\dag\subseteq \mathrm{G}_{Q_1}\Big)\geq \mathbf{P} \bigg(\mathcal{G}^\dag\subseteq \Big( \mcup_{\tau=1}^{|\mathcal{E}^\dag|}\mathrm{G}_{W(\zeta_\tau)}\Big)\bigg) \geq \Big(\frac{q_\ast }{n}\Big)^{|\mathcal{E}^\dag|}
\end{align*}
according to Lemma \ref{lem5}.

We continue to define $Q_m=W(\zeta_{m|\mathcal{E}^\dag|})\cdots  W(\zeta_{(m-1)|\mathcal{E}^\dag|+1})$ for $m=2,3,\dots$, and similarly
\begin{align*}
\mathbf{P} \Big(\mathcal{G}^\dag\subseteq \mathrm{G}_{Q_m}\Big)\geq \Big(\frac{q_\ast }{n}\Big)^{|\mathcal{E}^\dag|}
\end{align*}
for all $m$. Because $\mathcal{G}^\dag$ is quasi-strongly connected,  Lemma \ref{lem6}  yields
\begin{equation}\label{66}
\mathbf{P}\Big(\lambda \big(Q_{n-1}\cdots Q_1\big)<1\Big)\geq \Big(\frac{q_\ast }{n}\Big)^{(n-1)|\mathcal{E}^\dag|}.
\end{equation}
Moreover, $Q_{n-1}\cdots Q_1$ represents a product of $(n-1)|\mathcal{E}^\dag|$ stochastic matrices, each of which satisfies the weights rule. Therefore, there holds
\begin{equation}\label{67}
[Q_{n-1}\cdots Q_1]_{ij}\geq \eta^{(n-1)|\mathcal{E}^\dag|}
\end{equation}
for every  nonzero entry $[Q_{n-1}\cdots Q_1]_{ij}$ of $Q_{n-1}\cdots Q_1$.

We see from (\ref{66}) and (\ref{67}) that
\begin{equation*}
\mathbf{P}\Big(\lambda \big(Q_{n-1}\cdots Q_1\big)\leq1-\eta^{(n-1)|\mathcal{E}^\dag|}\Big)\geq \Big(\frac{q_\ast }{n}\Big)^{(n-1)|\mathcal{E}^\dag|}.
\end{equation*}
Denoting $U_m=Q_{m(n-1)}\cdots Q_{(m-1)(n-1)+1}$ for $m=1,2, \dots$, we can now further conclude for all $ m=1,2,\dots$ that
\begin{equation*}
\mathbf{P}\Big(\lambda \big(U_m\big)\leq 1-\eta^{(n-1)|\mathcal{E}^\dag|}\Big)\geq \Big(\frac{q_\ast }{n}\Big)^{(n-1)|\mathcal{E}^\dag|}.
\end{equation*}

 Thus, based on Lemma \ref{lem0}, we have
\begin{equation}\label{68}
\lim_{m\rightarrow \infty}\mathbf{E} \Big[\delta \big (U_m\cdots U_1\big)\Big]\leq \lim_{m\rightarrow \infty} \mathbf{E} \Big[\prod_{\tau=1}^{m} \lambda \big(U_m\big) \Big]= 0,
\end{equation}
Then from Fatou's Lemma we have
\begin{equation*}
\mathbf{P}\Big( \lim_{m\rightarrow \infty}\delta\big (U_m\cdots U_1\big)=0\Big)=1.
\end{equation*}
which leads to
$$
\mathbf{P} \Big(\lim_{k\rightarrow\infty}\delta \big(W(k)\cdots W(0)\big)=0\Big) =1
$$
 since $W(k)=I_n$ for any $k\notin\{\zeta_1,\zeta_2,\dots\}$ ($I_n$ is the identity matrix).  Hence, we have proved global a.s. consensus for Algorithm (\ref{9}).
\subsection{Proof of Theorem \ref{theorem-ARC-independent}: Computation Time}
In this subsection, we establish the upper bound of $\mathscr{T}_{\rm com}(\epsilon)$ given in Theorem \ref{theorem-ARC-independent}.

Denote $\Theta_k=W(k-1)\dots W(0)$. For all $i,j$ and $\tau$, there holds
\begin{align*}
\big|x_{i}(k)-x_j(k)\big|&=\Big|\sum_{\alpha=1}^n [\Theta_k]_{i\alpha }x_{\alpha }(0)-\sum_{\alpha=1}^n [\Theta_k]_{j\alpha }x_{\alpha }(0)\Big|\nonumber\\
&=\Big|\sum_{\alpha=1}^n [\Theta_k]_{i\alpha }\big(x_{\alpha }(0)-x_\tau(0)\big)-\sum_{\alpha=1}^n [\Theta_k]_{j\alpha }\big(x_{\alpha }(0)-x_\tau(0)\big)\Big|\nonumber\\
&\leq\sum_{\alpha=1}^n\big| [\Theta_k]_{i\alpha }- [\Theta_k]_{j\alpha }\big|\cdot \max_{\alpha}\big|x_{\alpha }(0)-x_\tau(0)\big|\nonumber\\
&\leq n \delta \big(\Theta_k\big)\cdot \max_{\alpha}\big|x_{\alpha }(0)-x_\tau(0)\big|.
\end{align*}
Therefore, we obtain that for all $k\geq1$,
\begin{align}\label{a4}
\mathcal{H}(k)\leq n \delta \big(\Theta_k\big)\mathcal{H}(0).
\end{align}

Introduce
$$
\xi_k:=\sum_{i=0}^{k-1} \Psi_i,
$$
for  $k=0,1,\dots$. Then there holds
$$
\xi_k=\max\big\{m: \zeta_m \leq k-1\big\}.
$$

Denote $E_0=(n-1)|\mathcal{E}^\dag|$. Then according to Lemma \ref{lem0} and applying Markov's Inequality, (\ref{a4}) implies
\begin{align}\label{a6}
\mathbf{P}\Big(\frac{\mathcal{H}(k)}{\mathcal{H}(0)} \geq\epsilon \Big)&\leq\mathbf{P}\Big(\delta \big(\Theta_k\big) \geq \frac{\epsilon}{n} \Big)\nonumber\\
&\leq \frac{n}{\epsilon} \mathbf{E}\Big[ \delta \big(\Theta_k\big) \Big]\nonumber\\
&\leq  \frac{n}{\epsilon} \mathbf{E}\big[ \lambda_{\lfloor \frac{\xi_k}{E_0}\rfloor}\cdots \lambda_1\big],
\end{align}
where $E_0=(n-1)|\mathcal{E}^\dag|$,  and by definition $\lambda_m=\lambda  (U_m)$ with $U_m$ introduced in the previous subsection. Here  $\lfloor z \rfloor$ represents the largest integer no greater than $z$.

From (\ref{66}) we conclude that
\begin{equation*}
\mathbf{E} \big[ \lambda_m \big] \leq 1- \Big(\frac{ \eta q_\ast}{n}\Big)^{E_0}
\end{equation*}
for all $m$. This further yields
\begin{align}\label{a5}
 \mathbf{E}\Big[ \lambda_{\lfloor \frac{\xi_k}{E_0}\rfloor}\dots \lambda_1\Big]&=\mathbf{E}\Big[\mathbf{E}\big[ \lambda_{\lfloor \frac{\xi_k}{E_0}\rfloor}\dots \lambda_1\big|\xi_k\big]\Big]\nonumber\\
 &\leq \Big(1- \Big(\frac{ \eta q_\ast}{n}\Big)^{E_0}\Big)^{\mathbf{E} \big[ \lfloor \frac{\xi_k}{E_0} \rfloor\big]}\nonumber\\
 &\leq \Big(1- \Big(\frac{ \eta q_\ast}{n}\Big)^{E_0}\Big)^{{\sum_{i=0}^{k-1}\mathbf{E} [\Psi_i]}/{E_0} -1}\nonumber\\
 &= \Big(1- \Big(\frac{ \eta q_\ast}{n}\Big)^{E_0}\Big)^{{\sum_{i=0}^{k-1}[1-(1-P_i)^n]}/{E_0}-1}.
\end{align}

Combining (\ref{a6}) and (\ref{a5}),  we obtain
\begin{align*}
\mathbf{P}\Big(\frac{\mathcal{H}(k)}{\mathcal{H}(0)} \geq\epsilon \Big)\leq  \frac{n}{\epsilon} \Big(1- \Big(\frac{ \eta q_\ast}{n}\Big)^{E_0}\Big)^{{\sum_{i=0}^{k-1}[1-(1-P_i)^n]/E_0}-1},
\end{align*}
based on which   the desired upper bound of $\mathscr{T}_{\rm com}(\epsilon)$ follows immediately  from some simple algebra.

\section{Examples, Mixing, and Gossiping}\label{Sec:Generalization}
In this section, we first investigate a few basic  random graph models  from the literature discussing  the applicability of the  previously established results.  Next,  we   extend the concepts  of connectivity-independence and arc-independence  to $\ast$-{\it mixing} of random graphs, and establish the corresponding a.s. consensus convergence result. An immediate
 advantage of this  extension is that the majority of random graph models is covered by $\ast$-mixing random graphs. Finally, we look further into random gossiping algorithms and present results on a.s. finite-time
 consensus for this important class of random algorithms.

\subsection{Examples}\label{Subsec:Examples}

We discuss a few random graph examples and show the applicability of the connectivity and arc independence. Apparently all independent random graph processes are immediately connectivity  independent, but not necessarily arc independent.

Let $\mathsf{K}_n$ denote the complete digraph with $n$ nodes,
i.e., $(i,j)\in \mathsf{K}_n$ for all $i\neq j \in \mathrm{V}$.

\medskip

\noindent {\bf Example 1} [Erd\H{o}s--R\'{e}nyi]. The random graph process $\langle \mathrm{G}_k\rangle$ is a sequence of i.i.d. bidirectional or directed
Erd\H{o}s--R\'{e}nyi random graphs. A bidirectional Erd\H{o}s--R\'{e}nyi graph is a random graph over node set $\mathrm{V}$ such that independently with a probability $p$,
 there is a bidirectional  edge between every pair of nodes in  $\mathrm{V}$. A directed Erd\H{o}s--R\'{e}nyi graph is defined in that independently with a probability $p$, there is
  a directed arc $(i,j)$ for every ordered pair  $(i,j)$ of nodes in $\mathrm{V}$. Consensus dynamics over Erd\H{o}s--R\'{e}nyi random graphs were studied in \cite{hatano,wu}.

Both  the directed and bidirectional
 Erd\H{o}s--R\'{e}nyi graph processes  are arc-independent with respect to the basic graph $\mathsf{K}_n$. Note that arc appearance in a fixed bidirectional Erd\H{o}s--R\'{e}nyi random graph ${G}_k$ is not independent, since $(i,j)\in \mathrm{E}_k$ implies $(j,i)\in \mathrm{E}_k$.

 \medskip

\noindent {\bf Example 2} [Random Link Failure]. Let $\mathcal{G}^{\rm u}=(\mathrm{V}, \mathcal{E}^{\rm u})$ be an underlying digraph representing fundamentally who can possibly
talk to whom in the network. Independently at different time instants and among different arcs in  $ \mathcal{E}^{\rm u}$, $(i,j)\in \mathrm{E}_k$ with a given probability $p_{ij}>0$.
In this way we obtain  an i.i.d. random graph process $\langle \mathrm{G}_k\rangle$. This model describes independent random failure in node communications, and consensus algorithm with random
link failure is studied in \cite{moura1,moura3,bamieh,fagnani2}. Then  we see $\langle \mathrm{G}_k\rangle$ given by this
random link failure process is arc-independent with respect to the basic graph $\mathcal{G}^{\rm u}$.

%

 \medskip

\noindent {\bf Example 3} [Markovian Random Graph].   The random graph process $\langle \mathrm{G}_k\rangle$ is generated by a  Markov chain whose  state space is (or,  a subset of)
 $\mathscr{G}$. Such examples were studied in \cite{boyd1,markov}. Since Markovian random graphs  introduce local dependence in time, in general it will be difficult for
 $\langle \mathrm{G}_k\rangle$
 to be either connectivity or arc independent. However, if the Markov chain is defined on the subset of quasi-strongly connected digraphs in $\mathscr{G}$,
 then connectivity-independence holds trivially for $\langle \mathrm{G}_k\rangle$.

 \medskip

\noindent {\bf Example 4} [Random Gossiping].  Independently at each time step,  a pair of nodes are randomly selected for averaging their states, called {\it gossiping}.
 Comprehensive analysis for random gossiping algorithms is established in \cite{boyd}, and gossiping algorithm also serves as a model for social belief propagation \cite{daron,como}.
 An independent random gossiping process $\langle \mathrm{G}_k\rangle$ is  arc-independent and trivially connectivity-independent (the probability of $\mathrm{G}_k$ being quasi-strongly
 connected equals  zero if $n>2$, and  one if $n=2$). We can  consider the random  joint graphs $\mathrm{G}\big([mT, (m+1)T-1]\big)$ over time intervals $[mT, (m+1)T-1]$
 for $m\in \mathbb{N}$, in which  $\big\langle \mathrm{G}\big([mT, (m+1)T-1]\big)\big \rangle$ is both  connectivity and arc independent, and   for  sufficiently large integer $T$ the condition of Proposition  \ref{prop1} will  be satisfied.

\subsection{Connectivity-Mixing and Arc-Mixing Random Graphs}\label{Subsec:Mixing}
In the previous subsection, we have illustrated that the concepts of connectivity/arc independence can be applied to
many  independent random graph models, but become restrictive with non-independent, e.g., Markovian random graph processes. We now propose an alternative method from {\it mixing} theory, which,
inherits the spirit of a network version of Borel-Cantelli lemma in the presence of connectivity/arc independence (and many our analysis techniques smoothly apply), however allows for possible
weakening of the independence assumption.

\subsubsection{Arc/Connectivity $*$-Mixing}

First we recall some basic concepts and results from mixing of random processes (cf., \cite{mixing} for a comprehensive survey).
Let $\langle X_k \rangle $ be a sequence of random variables, where each $X_k$ is taken from a common  probability space $(\Omega, \mathcal{F}, \mathbb{P})$. For the ease of presentation we
continue to use $(\Omega, \mathcal{F}, \mathbb{P})$ to denote the product probability space where $( X_k, k\in \mathbb{N})$ lies in, the existence of which
is guaranteed by the  Kolmogorov's extension theorem \cite{Durr}. The $\sigma$-algebra  generated by $X_a,\cdots, X_b$ is denoted as $\mathcal{F}_a^b$ for $0\leq a \leq b \leq \infty$.
The tail $\sigma$-algebra of $\langle X_k \rangle $, denoted $\mathcal{T}$, is defined by $\mathcal{T}=\mcap_{m=0}^\infty \mathcal{F}_m^\infty$. Events belonging to $\mathcal{T}$
are called {\it tail events}. For a sequence of events $\langle \mathpzc{A}_k \rangle$ with $\mathpzc{A}_k \in\mathcal{F}$, we define $\mathrm{I}_{ \mathpzc{A}_k}$ as their indicators,
i.e.,  $\mathrm{I}_{ \mathpzc{A}_k}(\omega)=1$ if $\omega\in  \mathpzc{A}_k$ and $\mathrm{I}_{ \mathpzc{A}_k}(\omega)=0$ otherwise for all $k\in\mathbb{N}$.
Then $\langle \mathrm{I}_{ \mathpzc{A}_k}\rangle$ is a sequence of random variables over  $(\Omega, \mathcal{F}, \mathbb{P})$.
We quote the following definition from \cite{starmixing1963}\footnote{There are various types  of mixing \cite{mixing}. The $*$-mixing originated in \cite{starmixing1963} we quote
in this paper corresponds to the $\psi$-mixing in \cite{mixing}. }.

\begin{definition}\label{mixingdefinition}
A sequence of random variables  $\langle X_k \rangle $ is called {\it $*$-mixing} if there exists a non-increasing sequence of real numbers $\langle f_m \rangle$ with $\lim_{m\to \infty} f_m=0$, such that
there holds
\begin{align*}
\big|\mathbb{P} (\mathpzc{A} \mcap \mathpzc{B} ) - \mathbb{P}(\mathpzc{A})\mathbb{P}(\mathpzc{B})\big|\leq f_m\mathbb{P}(\mathpzc{A})\mathbb{P}(\mathpzc{B}),
\ \mathpzc{A} \in \mathcal{F}_0^t, \mathpzc{B}\in \mathcal{F}_{t+m}^\infty
\end{align*}
for all   $t\in\mathbb{N}$. A sequence of events $\langle \mathpzc{A}_k \rangle$ is {\it $*$-mixing} if
$\langle \mathrm{I}_{ \mathpzc{A}_k}\rangle$ is  {\it $*$-mixing}.
\end{definition}

 Associated with $\langle \mathrm{G}_k \rangle$, we let
$\langle \mathpzc{C}_k \rangle$ be the sequence of connectivity events defined by $$
\mathpzc{C}_k:=\Big\{\omega: \mathrm{G}_k(\omega)\ \mbox{is quasi-strongly connected}\Big\},\ \ k\in\mathbb{N}.
$$
Similarly, take  a deterministic digraph $\mathcal{G}^\dag=(\mathrm{V}, \mathcal{E}^\dag)$ with
 $\mathcal{E}^\dag=\big\{(i_\tau,j_\tau):\tau=1,\dots, |\mathcal{E}^\dag|\big\}$.
Denote $$
\mathpzc{E}_k(\tau):=\big\{(i_\tau,j_\tau)\in\mathrm {G}_k\big\}
$$ for $\tau=1,\dots, |\mathcal{E}^\dag|$ and $k\in \mathbb{N}$. We can now introduce the following concepts of $*$-mixing for the random graph process $\langle \mathrm{G}_k \rangle$.

\begin{definition} A random graph process $\langle \mathrm{G}_k \rangle $ is termed

(i) connectivity  {\it $*$-mixing} if $\langle \mathpzc{C}_k \rangle$ is  {\it $*$-mixing};

(ii) arc {\it $*$-mixing} with respect to $\mathcal{G}^\dag=(\mathrm{V}, \mathcal{E}^\dag)$ if  $\langle \mathpzc{E}_k(\tau_k)\rangle $ is  {\it $*$-mixing} for any deterministic sequence  $\langle \tau_k \rangle $ with each $\tau_k$ taking value from $\{1,\dots, |\mathcal{E}^\dag|\}$.

\end{definition}

\subsubsection{Consensus for $*$-Mixing Random Graphs}
We now present our main consensus convergence result for random graph processes that are connectivity or arc $*$-mixing.

\begin{theorem}\label{thmmixing} Let  $\langle \mathrm{G}_k \rangle $ be a random graph process equipped with a probability measure $\mathbb{P}$.

  (i) Assume that $P_{k+1}\leq P_k$. Suppose there exist  an integer $B\geq 1$ and a constant $0<q\leq 1$ such that
\begin{itemize}
\item  $ \Big\langle\mathrm {G}\big([mB, (m+1)B-1]\big)\Big\rangle$ is connectivity  {\it $*$-mixing};

\item    $ \mathbb{P} \big( \mathrm {G}\big([mB, (m+1)B-1]\big)\mbox{ is quasi-strongly connected}\big) \geq q$ for all $k$.
\end{itemize}
Then Algorithm  (\ref{9}) achieves global a.s. consensus  if $\sum_{s=0}^\infty {P}_k^{n-1} =\infty$.

  (ii)  Assume that  $\langle \mathrm{G}_k \rangle $ is  arc {\it $*$-mixing} with respect to $\mathcal{G}^\dag$. Suppose  there is a constant $0<q\leq 1$ such that
  $ \mathbb{P} \big( (i,j)\in \mathrm{G}_k \big) \geq q$ for all $(i,j)\in \mathcal{E}^\dag$ and all $k\in \mathbb{N}$. Then Algorithm  (\ref{9}) achieves global a.s. consensus  if and only if
   $\sum_{s=0}^\infty {P}_k =\infty$.

\end{theorem}

This theorem is proved following the same lines as the  proofs of Proposition \ref{prop1} and Theorem~\ref{theorem-ARC-independent} for connectivity-independent and arc-independent graphs,
where in the argument whenever necessary,  we replace the standard  Borel-Cantelli Lemma for a sequence of independent events
 with the corresponding Borel-Cantelli Lemma for $*$-mixing sequence of events established in \cite{starmixing1963}.    The proof of Theorem \ref{thmmixing} is in Appendix B.

 On the other hand,
 the bounds of   $\mathscr{T}_{\rm com}(\epsilon)$ established in Proposition \ref{prop1} and Theorem~\ref{theorem-ARC-independent} no longer apply to the
 $*$-mixing random graphs. The reason is that the recursion of $\mathbf{E}\big[\mathcal{H}(k)\big]$ with arc/connectivity independence
 fails to stand under $*$-mixing, and therefore we cannot obtain the bounds of $\mathscr{T}_{\rm com}(\epsilon)$ from the  Markov's inequality. We however would like to point out that
 making use of the Strong Law of Large Numbers for $*$-mixing sequence of random variables  (cf., Theorem 1, \cite{starmixing1963}), one can still build upper bounds for
   $\mathscr{T}_{\rm com}(\epsilon)$ under the $*$-mixing conditions in Theorem \ref{thmmixing}. Apparently in this case
    $\mathscr{T}_{\rm com}(\epsilon)$ relies  not only on the presence of  $*$-mixing, but also on the speed of  $*$-mixing, i.e., the $f_m$ in Definition \ref{mixingdefinition}.

\subsubsection{An Illustrative Example}
It is known that  for a strictly stationary, finite-state Markov chain, it is   $*$-mixing if and only if it is irreducible and aperiodic (cf., Theorem 3.1, \cite{mixing}). Then it becomes
immediate that Theorem \ref{thmmixing} is applicable to a large class of Markovian random graph processes $\langle \mathrm{G}_k \rangle$. We give an example below.

 \medskip

\noindent {\bf Example 5} [Random Walk on Graphs].  We define the random graph process  $\langle \mathrm{G}_k\rangle$ as follows.
Let $\mathcal{G}^{\rm I}=(\mathrm{V}, \mathcal{E}^{\rm I})$ be a  strongly connected  digraph for node interactions.  Let $\langle s_k\rangle$ be an  irreducible and aperiodic
 Markov chain with state space $\mathrm{V}$ and state transition matrix $P=[P_{ij}]$. Then $\langle s_k\rangle$ gives a random walk on the graph $\mathcal{G}_{P}$, where
 $\mathcal{G}_{P}$ is the induced graph of matrix $P$. The random graph $\mathrm{G}_k$ is defined as $\mathrm{E}_k= \{j:(j,i)\in \mathcal{E}^{\rm I}\}$ if $s_k=i$. Algorithms of this type
 have been used in networked sub-gradient algorithms for convex optimization \cite{boyd1}.

 Let $\langle s_k\rangle$ be initialized at its invariant distribution and take $B_0=|\mathcal{E}^{\rm I}|$. It is then straightforward to conclude that
 $ \Big\langle\mathrm {G}\big([mB_0, (m+1)B_0-1]\big)\Big\rangle$ is connectivity  {\it $*$-mixing} and there exists $q_0>0$ such that
  $ \mathbb{P} \big( \mathrm {G}\big([mB, (m+1)B-1]\big)\mbox{ is quasi-strongly connected}\big) \geq q_0$ for all $k$. Moreover,
  $ \Big\langle\mathrm {G}\big([mB_0, (m+1)B_0-1]\big)\Big\rangle$ is also arc  {\it $*$-mixing} with respect to $\mathcal{G}^{\rm I}$ and there also exists $q'_0>0$ such that
  $ \mathbb{P} \big( (i,j)\in \mathrm {E}\big([mB, (m+1)B-1]\big)\big) \geq q'_0$ for all $(i,j)\in \mathcal{E}^{\rm I}$ and for all $k$. Therefore, Theorem \ref{thmmixing}
  is directly applicable.

\subsection{Random Gossiping}\label{Subsec:Gossiping}
Random gossiping is an important model in engineering (computer-to-computer, wireless communication) \cite{boyd} and social networks \cite{daron}. The applicability of the established results to random gossiping algorithms has been discussed in Example 4. In standard random gossip algorithms, interactions always happen pairwise and the two interacting nodes average their states during their interaction.
This simple nature  of random gossiping certainly admits  deeper results beyond the previous discussions. In this subsection, we make a further investigation to random gossiping
algorithms.

\subsubsection{Gossiping Algorithms}
Let $S=[S_{ij}]$ be an $n\times n$ stochastic matrix. A gossip algorithm is given by a node pair selection process \cite{boyd,daron}.
\begin{definition}
Independently at each time $k\geq0$, (i)  a node $i\in\mathrm{V}$ is drawn    with probability $1/n$;
(ii) the selected node $i$ picks the pair $\{i, j\}$ with probability $S_{ij}$.
\end{definition}

 This random pair selection process generates a sequence of i.i.d. random  graphs $\langle \mathrm{G}_k \rangle$, where
$$
\mathrm{E}_k=\big\{(i,j), (j,i) \big\}
$$
when  node pair $\{i, j\}$ is selected. When $\mathrm{E}_k=\big\{(i,j), (j,i) \big\}$, Algorithm  (\ref{9}) under this gossiping interaction reads as

{\em
\begin{itemize}
\item [(i)] $x_i(k+1)=
		x_i(k)/2+x_j(k)/2$ {with probability $P_k$}, and $x_i(k+1)= x_i(k)$ {with probability $1-P_k$};
\item [(ii)] Independent with node $i$, $x_j(k+1)=
		x_i(k)/2+x_j(k)/2$ {with probability $P_k$}, and $x_j(k+1)= x_j(k)$ {with probability $1-P_k$};
\item [(iii)] $x_j(k+1)=x_j(k)$ for all $j\notin\{i,j\}$.
\end{itemize}
}

This model generalizes the standard gossiping algorithm in  \cite{boyd} with $\langle P_k\rangle$ characterizing possible communication failure in every gossiping pairs. Let the induced graph of
$S$, $\mathcal{G}_S$, be quasi-strongly  connected. We can apply Theorem \ref{theorem-ARC-independent} and directly conclude that
 if and only if $\sum_{k=0}^\infty P_k =\infty$, the above random gossiping algorithm converges almost surely.
 On the other hand, considering the following random graph process
$$
\Big\langle \mathrm{G}\big([m|\mathcal{E}_S|, (m+1)|\mathcal{E}_S|-1]\big)\Big \rangle,
$$
there holds
$$
\mathbf{P}\Big( \mathrm{G}\big([m|\mathcal{E}_S|, (m+1)|\mathcal{E}_S|-1]\big) \mbox{is quasi-strongly connected}\Big)\geq (S_\ast/n)^{|\mathcal{E}_S|}
$$
with $S_\ast=\min \{S_{ij}+S_{ji}: S_{ij}+S_{ji}>0\}$. Therefore,  Theorem \ref{theorem-connectivity-independent} is also applicable to random gossiping process.

This structure of $\mathrm{E}_k$ in this random gossiping process however allows us to go beyond the asymptotic consensus analysis.
We define a.s. finite-time  consensus in the following.
\begin{definition}
Global finite-time a.s. consensus is achieved for Algorithm (\ref{9}) if  for any initial value $x^0$, there
exits  a random time $K_{x^0}(\omega)\in \mathbb{N}$ which is  a.s. finite such that
$\mathbf{P}_{x^0}\big(\mathcal{H}(K_{x^0})=0 \big)=1$. The finite-time computation time $\mathscr{T}^{\rm f}_{\rm com}$, is defined as
$$
\mathscr{T}_{\rm com}^{\rm f}=\sup_{x^0\in \mathbb{R}^n} \min_{k}\Big\{k\in \mathbb{N}: \mathcal{H}(k)=0\Big\}.
$$
\end{definition}

Note that $\mathscr{T}_{\rm com}^{\rm f}$ is by definition a random variable, in contrast to  $\mathscr{T}_{\rm com}(\epsilon)$ which is a constant.
\subsubsection{Almost Sure Finite-time Convergence}

%
%

%

For  finite-time convergence of random gossiping process, we present the following result.

\begin{theorem}\label{Prop-Gossip-FiniteTime} Let $\langle \mathrm{G}_k \rangle$ be a random gossiping process defined by stochastic matrix $S=[S_{ij}]$. Suppose  $\mathcal{G}_S= \mathsf{K}_n$. Suppose $n=2^m+r$ with $0\leq r< 2^m$.

(i). Let  $0<P_\ast<1/2$ be a constant such that $P_\ast \leq P_k \leq 1-P_\ast$ for all $k$ and suppose  $\langle P_k \rangle $ is monotonic.  Then global a.s. finite-time consensus is achieved if $\sum_{k=0}^\infty \big[P_k(1-P_k)\big]^{ 2 N_0} =\infty$ with $N_0=r+m(n+r)/2$. Moreover,
 in this case there holds  $$
 \mathbf{E} \big[\mathscr{T}_{\rm com}^{\rm f} \big ]\leq  N_0 \Big(\frac{n}{ P_\ast^2 S_\ast}\Big)^{N_0}.
 $$

(ii). Let  $P_k=1$ for all $k$.  Then global a.s. finite-time consensus is achieved if and only if there is an integer $m>0$ such that $n=2^m$. In fact,
\begin{itemize}
\item[a)]  When $n=2^m$, there holds
$$
\mathbf{E} \big[\mathscr{T}_{\rm com}^{\rm f} \big ]\leq N(n/S_\ast)^N
$$ with $N={(n \log_2 n)}/{2}$;
\item[b)] When $n$ is not some power of two (i.e., $r>0$), we have $\mathbf{P}_{x^0}\big(\mathcal{H}(k)>0, k\in \mathbb{N} \big)=0$ for almost all $x^0\in \mathbb{R}^n$ under standard Lebesgue measure.
\end{itemize}
\end{theorem}

The proof of Theorem \ref{Prop-Gossip-FiniteTime} is in Appendix C. The proof is obtained by incorporating  the results on finite-time convergence of deterministic
gossip algorithms \cite{MTNS} with the Borel-Cantelli
Lemma for independent sequence of events. Therefore, Theorem \ref{Prop-Gossip-FiniteTime} continues to hold for a large class of Markovian gossiping processes (e.g.,
in Example 5, node $i$ only selects a node $j$ from its neighbors for gossiping when $s_k=i$.), based on
the same  $*$-mixing analysis as Theorem~\ref{thmmixing}.

It is worth emphasizing from Theorem \ref{Prop-Gossip-FiniteTime} that,  interestingly enough, the random update process  on one hand decelerates the asymptotic
convergence of the gossiping algorithm, but on the other hand  makes  finite time convergence in general possible (cf., without random node updates, i.e., $P_k=1$ for all $k$,
a.s. finite-time convergence exists only if $n$ is some power of two, which is a rather strong restriction to the network).

\section{Conclusions}\label{Sec:Conclusion}
This  paper investigated standard consensus algorithms coupled with randomized  individual node decision-making over stochastically time-varying graphs. Each node determined its update by a sequence of Bernoulli trials with time-varying probabilities.   The central aim of this work was to investigate the relation between the required level of independence of the graph process and the overall convergence. We consequently introduced    connectivity-independence and arc-independence for the random graph processes. In light of the Borel-Cantelli lemma, a universal  impossibility theorem showed that  a.s. consensus cannot be achieved unless the sum of the success probability sequence  diverges. Then a series of sufficiency conditions were given for the network to reach a global a.s. consensus under various connectivity assumptions. We established sharp threshold of almost sure  consensus obeying the zero-one law.    Convergence rates are established by lower and upper bounds of  the $\epsilon$-computation time. We also generalized the concepts of connectivity/arc independence to their analogues from the $*$-mixing point of view, which covers  the majority of random graph models in the literature such as Erd\H{o}s-R\'{e}nyi, gossiping, random walk, and Markovian random graphs. Under the $*$-mixing setting, our convergence  analysis continued  to stand  and almost sure consensus conditions were established. Finally, we further investigated  almost sure finite-time convergence of random gossiping algorithms.  Surprisingly,  the random node update process plays a key role in making a.s. finite-time convergence possible.

\medskip

\medskip

\medskip

\medskip

\noindent{\Large \bf Appendices}

\subsection*{Appendix A. Proof of Theorem \ref{theorem-Impossibility}}
The proof relies on a few preliminary lemmas. First of all, the following lemma is well-known (see Theorem 15.5, pp. 300, \cite{Rudin}).
\begin{lemma}\label{lem1}
Suppose $0\leq b_k<1$ for all $k$. Then $\sum_{k=0}^\infty b_k=\infty$ if and only if $\prod_{k=0}^{\infty}(1-b_k)=0$.
\end{lemma}

\begin{lemma}\label{lem8}
Suppose there exists a constant $a_\ast$ such that  $a_{ii}(k)\geq  a_\ast  >1/2$ for all $i$ and $k$. Then for all $k\geq0$, there always holds  $
\mathcal {H}(k+1)\geq\big(2 a_\ast  -1\big)\mathcal {H}(k)$.
\end{lemma}
{\it Proof.} Let  $x_{m}(k)=h(k)$ for some $m\in\mathrm{V}$. Then we have
\begin{align}
\sum_{j\in \mathrm{N}_{m}(k)}a_{mj}(k)x_j(k)
&=a_{mm}(k)x_{m}(k)+\sum_{j\in \mathrm{N}_{m}(k)\setminus \{m\}}a_{mj}(k)x_j(k)\nonumber\\
&\leq a_{mm}(k)h(k)+ \big(1-a_{mm}(k)\big)H(k)\nonumber\\
 &\leq a_\ast   h(k)+\big(1- a_\ast  \big)H(k), \nonumber
\end{align}
which implies
\begin{align}\label{t1}
h(k+1)\leq  a_\ast   h(k)+\big(1- a_\ast  \big)H(k).
\end{align}
A symmetric argument leads to
\begin{align}\label{t2}
H(k+1)\geq \big(1- a_\ast  \big) h(k)+ a_\ast  H(k).
\end{align}
Based on (\ref{t1}) and (\ref{t2}), we obtain
\begin{align}
\mathcal{H}(k+1)&=H(k+1)-h(k+1)\nonumber\\
&\geq \big(1- a_\ast  \big) h(k)+ a_\ast  H(k) -\big[ a_\ast   h(k)+\big(1- a_\ast  \big)H(k)\big]\nonumber\\
&\geq \big(2 a_\ast  -1\big)\mathcal{H}(k).
\end{align}
The desired conclusion follows.\hfill$\square$

Let  $\mathscr{S}_0$ be a set of stochastic matrices. Consider the following (deterministic) distributed averaging algorithms
\begin{align}\label{r10}
 x(k+1)=W_kx(k),
 \end{align}
 where each $W_k\in \mathscr{S}_0$ for all $k$.  We define
 $$
\mathscr{Z}_0\doteq  \Big\{z\in \mathbb{R}^n: \ \exists W_0,\dots,W_s\in \mathscr{S}_0, s\geq 0\ \ {\rm s.t.}\  W_s\cdots W_0z\in {\rm span}\{\mathbf{1}\} \Big\}.
 $$

Let ${\mu}(\cdot)$ represent the standard Lebesgue measure on $\mathbb{R}^n$. The following lemma from \cite{MTNS} holds for the finite-time convergence of averaging algorithm (\ref{r10}).
\begin{lemma}\label{lemallnothing}
Suppose $\mathscr{S}_0$ is a set with at most countable elements.  Then either $\mathscr{Z}_0=\mathbb{R}^n$ or  ${\mu}(\mathscr{Z}_0)=0$. In fact, if $\mathscr{Z}_0\neq\mathbb{R}^n$, then $\mathscr{Z}_0$ is a union of at most countably many linear spaces whose dimensions are no larger than $n-1$.
\end{lemma}

We are now ready to present the detailed proof of Theorem \ref{theorem-Impossibility}.

\medskip

\noindent {\it Proof of (i)}.   Let $\mathcal{H}(0)>0$. Then  there exist at least
 two nodes, say $i$ and $j$ with different initial conditions. Considering
the probability that both node $i$ and node $j$ remain constant: From  Algorithm (\ref{9}), we have
\begin{align}
\mathbf{P}\Big(x_i(k+1)=x_i(k), \ {\rm and}\ x_j(k+1)=x_j(k)\ k=0,1,\dots\Big)\geq \Big[ \prod_{k=0}^{\infty}(1-P_k) \Big]^2 \doteq r_0^2,\nonumber
\end{align}
 if $\sum_{k=0}^\infty P_k<\infty$, where $0<r_0=\prod_{k=0}^{\infty}(1-P_k)<1$  according to Lemma \ref{lem1}. The  necessity  claim holds.

Next,  let  $\mathcal{H}(0)>0$ and denote by $i$ and $j$ two nodes satisfying $x_i(0)=h(0)$ and $x_j(0)=H(0)$, respectively.  Notice that when  both of them
never make a state update,  they  will remain at all times with minimum
(resp. maximum) states, because the other nodes cannot go further by making convex
combinations. Hence,
$$
\mathds{P}\Big({\mathcal{H}(k)}> \epsilon {\mathcal{H}(0)} \Big) \geq P\Big({\mathcal{H}(k)}\geq {\mathcal{H}(0)}\Big) \geq \prod_{h=0}^{k-1} (1-P_h)^2.
$$
Then the lower bound of the $\epsilon$-computation  can be easily obtained.

\medskip

\noindent {\it Proof of (ii)}. We rewrite Algorithm (\ref{9}) into
\begin{align}
x(k+1)=A_k x_k
\end{align}
with $A_k=[a_{ij}(k)]$ being a random matrix taking value from the set of stochastic matrices. Now that  $\big\{a_{ij}(k):i,j\in\mathrm{V}, k\in\mathbb{N}\big\}$ contains at most countably many elements, applying Lemma \ref{lemallnothing} we conclude that one of the following two cases must happen:
\begin{itemize}
\item[$a^\dag$)] There exists a constant $K\geq 0$ under which  a sample path $A_k(\omega)$ satisfies  $$
{\rm rank}\big(A_K(\omega)\dots A_0(\omega)\big)=1.
$$
This implies that $\mathcal{H}(K)(\omega)=0$ for all initial value $x^0$.
\item  [$b^\dag$)]  For almost all initial value $x^0$ and all $\omega$, $\mathcal{H}(k)(\omega)>0$ for all $k\in\mathbb{N}$.
\end{itemize}

The  First Borel-Cantelli Lemma \cite{Durr} ensures that every node updates its state only for some finite times when $\sum_{k=0}^\infty P_k<\infty$. Therefore, it becomes immediate to see that the cases $a^\dag$) and $b^\dag$) correspond to cases $a)$ and $b)$, respectively.

\medskip

\noindent {\it Proof of (iii)}.  Let $\mathcal{H}(0)>0$. There holds
\begin{align}\label{t9}
\sum_{k=0}^\infty P_k<\infty\  &\Leftrightarrow\  \prod_{k=0}^\infty (1-P_k)>0\Leftrightarrow\  \prod_{k=0}^\infty (1-P_k)^n>0\Leftrightarrow\  \sum_{k=0}^\infty \big(1-(1-P_k)^n\big)<\infty,
\end{align}
where the last equivalence is obtained by taking $b_k=1-(1-P_k)^n$ in Lemma \ref{lem1}. From the definition of Algorithm (\ref{9})  we see that
\begin{align}\label{t6}
\mathbf{P}\Big( \mathcal{H}(k+1)< \mathcal{H}(k) \Big)\leq \mathbf{P}\Big( \mbox{at least one node takes averaging at time $k$}\Big)= 1- (1-P_k)^n.
\end{align}

In light of  Lemma \ref{lem8}, and  applying the First Borel-Cantelli Lemma \cite{Durr} on (\ref{t6}), it follows immediately that
\begin{align}
\mathbf{P}\Big( \lim_{k\rightarrow \infty} \mathcal{H}(k)=0 \Big)\leq \mathbf{P}\Big(  \mathcal{H}(k+1)< \mathcal{H}(k) \mbox{\ for infinitely many $k$}\Big)=0.
\end{align}
The desired conclusion follows.

We have now completed the proof. \hfill$\square$

\subsection*{Appendix B. Proof of Theorem \ref{thmmixing}}

The proof is based on the following  Borel-Cantelli lemma for $*$-mixing sequence of events.
\begin{lemma}\label{Lem-mixingborelcantelli} (Lemma 6, \cite{starmixing1963})
Let  $\langle \mathpzc{A}_k \rangle$ be a sequence of events which is  {\it $*$-mixing}.
Then $\sum_{k=0}^\infty \mathbb{P}(\mathpzc{A}_k)=\infty$ implies $\mathbb{P}(\limsup_{k \to \infty} \mathpzc{A}_k)=1$.
\end{lemma}

\noindent (i). Let $k_s=s(n-1)^2 B$. We introduce the following sequence of events $\langle \mathfrak{E}_s \rangle $:
\begin{equation*}
\mathpzc{H}_s:= \Big\{\mathcal{H}(k_{s+1})\leq \Big(1-\frac{\eta^{(n-1)^2}}{2}\Big)\mathcal{H}(k_s)\Big\},  \ s=0,1,\dots.
\end{equation*}
We also define the sequence of event $\langle \mathpzc{B}_m\rangle $:
$$
\mathpzc{B}_m:=\Big\{\mathrm {G}\big([mB, (m+1)B-1]\big)\ \mbox{is quasi-strongly connected}\Big\},\ \ m=0,1,\dots.
$$

The proof of Proposition \ref{prop1} implies
\begin{align}\label{r20}
\mathpzc{H}_s \supseteq \mathpzc{Z}_s,
\end{align}
where by definition
\begin{align*}
\mathpzc{Z}_s:=\Big( \mcap_{m=0}^{(n-1)^2B-1} \mathpzc{B}_{s(n-1)^2 B+m}\Big) \mcap \Big\{\mbox{the $i_1,\dots,i_{n-1}\in\mathrm{V}$ in turn update their states}\Big\}.
\end{align*}
Noticing  the facts that $\langle\mathpzc{B}_m\rangle$ is $*$-mixing, and the  node updates are independent of  the graph process, node states and other nodes,
$\langle \mathpzc{Z}_s \rangle$ is also $*$-mixing. Moreover, with the monotonicity of $\langle P_k \rangle$, there holds
\begin{equation}\label{r21}
\mathbf{P} \big(\mathpzc{Z}_s\big) \geq  ({q}/{n})^{(n-1)^2} P_{{s(n-1)^2 B}}^{n-1}
\end{equation}
from the same argument as we obtain (\ref{s10}).

Therefore, we can apply  Lemma \ref{Lem-mixingborelcantelli} to $\langle \mathpzc{Z}_s \rangle$,  and conclude from (\ref{r20}) and (\ref{r21}) that
if $\sum_{k=0}^\infty {P}_k^{n-1} =\infty$, then
\begin{equation}\label{r22}
\mathbf{P} \big(\limsup_{s\to \infty} \mathpzc{H}_s\big)\geq \mathbf{P} \big(\limsup_{s\to \infty} \mathpzc{Z}_s\big) =1.
\end{equation}
Observing that the event $\limsup_{s\to \infty} \mathpzc{H}_s$ means the $\mathpzc{H}_s$ happen for infinitely many $s$, and that
$\mathcal{H}(k+1)\leq \mathcal{H}(k)$ always holds true, (\ref{r22}) immediately leads to global a.s. consensus for Algorithm (\ref{9}). We have
now completed the proof for connectivity $*$-mixing random graphs.

\medskip

\noindent (ii). Again, the proof is obtained by adapting  the proof of Theorem \ref{theorem-ARC-independent} to the case of arc $*$-mixing graphs. All we need  is to adopting
the above Borel-Cantelli argument  after (\ref{66}), instead of the original one using  Fatou's Lemma. The details are therefore omitted.

We have now completed the proof. \hfill$\square$

%
%
%
%
%
%

\medskip

\subsection*{Appendix C. Proof of Theorem \ref{Prop-Gossip-FiniteTime}}

The proof is built upon some preliminary results on deterministic finite-time gossiping  \cite{MTNS} (for a more comprehensive treatment we hereby refer to \cite{ToN}).  Introduce the following two sets of
stochastic matrices:
\begin{align*}
 \mathfrak{M}_n:=\Big\{ I_n-\frac{(e_i-e_j)(e_i-e_j)^T}{2}:\  i,j\in \mathrm{V} \Big\};\  \mathfrak{M}_n^\dag:=  \Big\{  I_n-\frac{e_i(e_i-e_j)^T}{2}:\  i,j\in\mathrm{V}\Big\}.
\end{align*}
The matrices in $ \mathfrak{M}_n$ and $ \mathfrak{M}_n^\dag$ represent the network state transition matrix for the realizations of  successful node pair interactions \cite{boyd}.

\medskip

\noindent (i). We recall the following conclusion established in \cite{MTNS}.

\begin{lemma}\label{lemma-gossip-asymme}
Suppose  $n=2^m+r$ with $m\geq 0$ and  $0\leq r<2^m$. Then there are $N_0=r+m(n+r)/2$ matrices $M_1,\cdots, M_{N_0}\in  \mathfrak{M}_n \mcup  \mathfrak{M}_n^\dag$ such that
${\rm rank} \Big(M_{N_0} \dots M_1\Big)=1$.
\end{lemma}

Now let  $M_1,\dots, M_{N_0}$ be the $N_0$ matrices defined in Lemma \ref{lemma-gossip-asymme} from
the set $\mathfrak{M}_n \mcup  \mathfrak{M}_n^\dag$ with  $${\rm rank} \Big(M_{N_0} \dots M_1\Big)=1.
$$ Each $M_k$ corresponds to a pair of nodes $\{i_k,j_k\}$.
The
probability of a given pair of nodes selected at time $k$ is no smaller than $S_\ast/n$ from the definition of the gossiping process. Depending on $M_k\in \mathfrak{M}_n $
or $M_k\in \mathfrak{M}_n^\dag$, there are two cases:
 $M_k$ is realized when both the two nodes $\{i_k,j_k\}$ successfully update their states; $M_k$ is realized when  only one node ($i_k$ or $j_k$) successfully updates its state. For
 either of the two cases we call the selected node pair  $\{i_k,j_k\}$ {\it realizes} $M_k$. Then at time $k$, $M_k$ is realized with probability either
 $P_k^2$ or $P_k(1-P_k)$ given that  $\{i_k,j_k\}$ is selected.

Consider the
following events:
$$
\mathpzc{E}_s:=\Big\{  (i_k,j_k)\in \mathrm{E}_{sN_0+k-1} \mbox{and $M_k$ is realized}, \ k=1,\dots, N_0\Big\},\ \ s\in \mathbb{N}.
$$
Note that  ${\rm rank} \Big(M_{N_0} \dots M_1\Big)$ implies that for all initial value $x^0$,
any $\mathpzc{E}_{s-1}$ happening will lead to $\mathcal{H}(sN_0)=0$.

Suppose that $\langle P_k \rangle $ is non-increasing and without loss of generality we let $P_0<1$. Then there holds for all $k$ that
$$
P_k(1-P_k)\geq (1-P_0)P_k \geq (1-P_0)P_k^2,
$$
and $\sum_{k=0}^\infty \big [P_k(1-P_k)\big]^{ 2N_0}=\infty$ leads to $\sum_{k=0}^\infty  P_k^{2 N_0}=\infty$. This gives us
\begin{align}
\mathbf{P}\big(\mathpzc{E}_{s-1}\big) \geq  \Big((1-P_0)(S_\ast/n)P_{sN_0}^2 \Big)^{N_0}.
\end{align}
Thus,  invoking the Borel-Cantelli lemma, a.s. finite-time consensus is achieved for Algorithm (\ref{1}) if $\sum_{k=0}^\infty\big [P_k(1-P_k)\big]^{ 2N_0}=\infty$.

The case when  $\langle P_k \rangle $ is non-decreasing holds from a symmetric argument, whose details  are therefore omitted.  This concludes the proof of a.s. finite-time convergence.

Next, letting     $P_\ast \leq P_k \leq 1-P_\ast$ for all $k$, we establish the upper bound for $\mathbf{E} \big[\mathscr{T}_{\rm com}^{\rm f} \big ]$.  Note that
there always holds
$$
\mathscr{T}_{\rm com}^{\rm f}\leq  N_0 \inf_s \Big\{s\geq 1: \ \mathpzc{E}_{s-1}\  \mbox{happens} \Big\}.
$$
As a result, we conclude that
\begin{align*}
\mathbf{E}\big[ \mathscr{T}_{\rm com}^{\rm f} \big] \leq \sum_{k=1}^\infty N_0 k (1-p)^{k-1} p= N_0 \sum_{k=0}^\infty (1-p)^k= N_0/p,
\end{align*}
where $p=( P_\ast^2 S_\ast/n)^{N_0}$ is a lower bound for the probability of $\mathpzc{E}_s$. This proves the proposed upper bound for $\mathbf{E}\big[ \mathscr{T}_{\rm com}^{\rm f} \big]$.

\medskip

\noindent (ii). The proof requires the following conclusion from \cite{MTNS}.
\begin{lemma}\label{lemsymmetric}
(i) If $n=2^m$, then there are  $N=\frac{n \log_2 n}{2}$ matrices  $M_1,\dots, M_{N}$
with each $M_i\in  \mathfrak{M}_n$ such that $M_{N}\cdots M_1=\mathbf{1}\mathbf{1}\T/n$.

(ii) If $n$ is not some power of two, then for almost all initial value $x^0$, there holds $$
M_{t}\cdots M_1 x^0 \notin {\rm span}\{\mathbf{1}\}
$$
for all $ M_1,\dots, M_t \in \mathfrak{M}_n $ and for all $t\in \mathbb{N}$.
\end{lemma}

\noindent {\it Proof of a).} Let  $n=2^m$ for some integer $m$. Let  $M_1,\dots, M_{N}$ be the $N=\frac{n \log_2 n}{2}$ matrices defined in Lemma \ref{lemsymmetric} from
the set $\mathfrak{M}_n$ with  $M_{N}\cdots M_1=\mathbf{1}\mathbf{1}\T/n$. Similarly, each $M_k$ corresponds to a pair of nodes $\{i_k,j_k\}$, and he
probability of a given pair of nodes selected at time $k$ is no smaller than $S_\ast/n$ from the definition of the gossiping process. Consider the
following events:
$$
\mathpzc{D}_s:=\Big\{  (i_k,j_k)\in \mathrm{E}_{sN+k-1}, \ k=1,\dots, N\Big\},\ \ s\in \mathbb{N}.
$$
Since $P_k=1$ for all $k$, the fact that $M_{N}\cdots M_1=\mathbf{1}\mathbf{1}\T/n$ implies that for all initial value $x^0$,
any $\mathpzc{D}_{s-1}$ happening will lead to $\mathcal{H}(sN)=0$. Therefore, a.s. finite-time consensus follows immediately from the Borel-Cantelli lemma. Moreover, there always holds
$$
\mathscr{T}_{\rm com}^{\rm f}\leq  N \inf_s \Big\{s\geq 1: \ \mathpzc{D}_{s-1}\  \mbox{happens} \Big\}.
$$
Similarly, with $p_\ast=(S_\ast/n)^N$ being  a lower bound for the probability of $\mathpzc{D}_s$,  we obtain
\begin{align*}
\mathbf{E}\big[ \mathscr{T}_{\rm com}^{\rm f} \big] \leq
N/p_\ast,
\end{align*}
 This proves the proposed upper bound for $\mathbf{E}\big[ \mathscr{T}_{\rm com}^{\rm f} \big]$.

\noindent {\it Proof of b).} The desired conclusion follows directly from Lemma \ref{lemsymmetric}.(ii), which indeed reveals the impossibility of finite-time consensus for every sample
path for almost all initial values.

We have now completed the proof of the proposition. \hfill$\square$

\medskip

\medskip

\medskip

\section*{Acknowledgements}
This work has been supported in part
by the Knut and Alice Wallenberg Foundation, the Swedish Research
Council, KTH SRA TNG, and by NICTA Ltd
and the Australian Research Council (ARC) under
DP-110100538 and DP-130103610.  A brief version of the current manuscript was presented  at the American Control Conference in Montreal, Canada, July 2012 \cite{ACC}.

\medskip

\medskip

{\noindent {\sc Guodong Shi and Brian D. O.  Anderson}} \\
{\noindent {\small  Research School of Engineering, College of Engineering and Computer Science, \\  The Australian National University, Canberra, ACT 0200, Australia.}}\\  {\small Email: } {\tt\small guodong.shi@anu.edu.au, brian.anderson@anu.edu.au}

\medskip

\medskip

\noindent {\sc Karl H. Johansson} \\
{\noindent {\small  ACCESS Linnaeus Centre,
   School of Electrical Engineering, \\
KTH Royal Institute of Technology, Stockholm 100 44, Sweden. }\\}
       {\small Email: {\tt\small
        kallej@kth.se}}

\end{document}